\documentclass[journal]{IEEEtran}
\usepackage{cite}
\usepackage[pdftex]{graphicx}
\graphicspath{{figures/}}
\usepackage{amsmath, amssymb}
\usepackage{array}
\usepackage{algorithm}
\usepackage{enumitem}
\usepackage{xspace}
\usepackage{makeidx} 
\usepackage{algpseudocode}
\usepackage[subrefformat=parens,labelformat=parens, caption=false,font=footnotesize]{subfig}
\usepackage{xcolor}
\usepackage{comment}
\usepackage{amsmath}
\usepackage{hyperref}

\DeclareMathOperator*{\argmax}{argmax}
\DeclareMathOperator*{\argmin}{argmin}
\DeclareMathOperator*{\synd}{syndrome}

\mathchardef\mhyphen="2D

\def\snr{\text{\rm SNR}\xspace}

\IEEEoverridecommandlockouts
\usepackage{fancyhdr}
\makeatletter
\renewcommand{\ALG@name}{Procedure }
\makeatother
\algnewcommand{\LineComment}[1]{\State \(\triangleright\) #1}

\begin{document}



\title{Decoding Short LDPC Codes via BP-RNN Diversity and Reliability-Based Post-Processing\thanks{This work was supported by the French Agence Nationale de la
Recherche (ANR), under grant number ANR-21-CE25-0006 (AI4CODE project).}}


\author{\IEEEauthorblockN{Joachim Rosseel\IEEEauthorrefmark{1}\IEEEauthorrefmark{2}, \quad Valérian Mannoni\IEEEauthorrefmark{1}, \quad Inbar Fijalkow\IEEEauthorrefmark{2}, \quad Valentin Savin\IEEEauthorrefmark{1}}\\
\IEEEauthorblockA{\IEEEauthorrefmark{1}Université Grenoble Alpes, CEA-Leti, F-38054 Grenoble, France\\
\IEEEauthorblockA{\IEEEauthorrefmark{2}ETIS, CY Cergy Paris Univ., ENSEA, CNRS, F-95000 Cergy-Pontoise, France\\ 
\{Joachim.Rosseel, Valerian.Mannoni, Valentin.Savin\}@cea.fr}, Inbar.Fijalkow@ensea.fr}}


%


\maketitle
\begin{abstract}
This paper investigates decoder diversity architectures for short low-density parity-check (LDPC) codes, based on recurrent neural network (RNN) models of the belief-propagation (BP) algorithm. We propose a new approach to achieve decoder diversity in the waterfall region, by specializing BP-RNN decoders to specific classes of errors, with absorbing set support.
We further combine our approach with an ordered statistics decoding (OSD) post-processing step, which effectively leverages the bit-error rate optimization deriving from the use of the binary cross-entropy loss function. We show that a single specialized BP-RNN decoder combines better than BP with the OSD post-processing step. Moreover, combining OSD post-processing with the diversity brought by the use of multiple BP-RNN decoders, provides an efficient way to bridge the gap to maximum likelihood decoding.
\end{abstract}


\section{Introduction}
\IEEEPARstart{L}{ast} years have witnessed an increased interest in research and practice of efficient error correcting codes for messages ranging from a few tens up to a few hundred bits, revived for instance by short-packet machine-to-machine communications, central to the emerging  Internet of Things  technology.
While important progress has been made in understanding the limits of coding at short block lengths~\cite{polyanskiy2010channel}, the design of efficient short codes and decoding algorithms still raises many challenges~\cite{cocskun2019efficient}.   

Low-density parity-check (LDPC) codes~\cite{gallager1963low} are a class of error correcting codes defined by sparse bipartite graphs~\cite{tanner1981recursive}. They are well-known for their excellent error correction performance at suitably large blocklengths, achieving near Shannon channel capacity performance under iterative belief propagation (BP) decoding~\cite{richardson2001design}. 
For codes defined by cycle-free bipartite graphs, BP decoding outputs  maximum a posteriori estimates of the coded bits \cite{ wiberg1996codes}. However, good codes are actually defined by graphs with cycles, in which case BP decoding is known to be sub-optimal.

First works on improving the iterative BP decoding performance actually focused on so-called high-density parity-check (HDPC) codes~\cite{kothiyal2005iterative, halford2006random, dimnik2009improved, hehn2010multiple}. These codes are defined by higher density bipartite graphs -- \emph{e.g.}, Bose-Chaudhuri-Hocquenghem (BCH) codes, Reed-Muller codes, binary Golay codes, etc. -- for which the standard iterative BP decoding usually yields very poor error correction performance.
In~\cite{kothiyal2005iterative}, an adaptive BP decoding approach was proposed, in which the decoding graph used by  BP  is updated at each iteration,  according to the output of a reliability-based decoding algorithm, such as the ordered statistics decoding (OSD)~\cite{fossorier1995osd}.
The random recurrent decoding  in~\cite{halford2006random,dimnik2009improved} and the multiple-bases BP decoding  in~\cite{hehn2010multiple} exploit a decoding diversity approach, in the form of different graph representations of the code, implying several BP decoders working either in  serial or in parallel.

More recently, deep neural networks have received significant interest for improving the decoding performance of short codes~\cite{nachmani2016learning, nachmani2018deep, xiao2020designing, xiao2021faid, buchberger2020pruning, doan2019neural}. A weighted BP decoding has been introduced in~\cite{nachmani2016learning}, where the weights are optimized using a neural network (NN).   
The topology of the NN mimics the BP decoding process, with unwrapped decoding iterations. The approach can be used with either  a feedforward (FF) or  a recurrent NN (RNN), the corresponding decoders being termed as BP-FF or BP-RNN. It has been shown in~\cite{nachmani2018deep} that the BP-RNN is able to outperform the standard BP decoder for short BCH codes, belonging to the class of HDPC codes. Subsequently, several variants of NN-based BP decoding  have been proposed in the literature. In~\cite{buchberger2020pruning},  a pruning method of irrelevant check-nodes in a neural BP model has been proposed, aimed at jointly optimizing the code construction and the decoding. The design of new decoding rules for finite-alphabet iterative decoders (FAIDs), based on a quantized NN model, was proposed in~\cite{xiao2020designing}. Moreover, a new training method of the quantized NN model was introduced in~\cite{xiao2021faid}, where training sets are constructed by sampling errors with trapping set support, to achieve decoding diversity for FAIDs on the binary symmetric channel. 
Resolving decoding failures due to trapping sets, by means of deep learning techniques, has also been recently investigated in~\cite{han2022deep}.

In this paper, we consider the BP-RNN decoding of short  LDPC codes. One critical difficulty faced by BP decoding in the finite blocklength regime is the presence of particular structures in the bipartite graph, which prevent the decoding algorithm from converging. Examples of such structures are trapping sets~\cite{richardson2003error} and absorbing sets~\cite{dolecek2009predicting}, and they are closely related to the pseudo-codewords~\cite{frey1998skewness} and near-codewords~\cite{mackay2003weaknesses} concepts.
Here, we shall focus on the absorbing set concept, introduced in~\cite{dolecek2009predicting} as a combinatorial object associated
with the bipartite graph, and  defined independently of the particular message-passing decoding or channel noise model. For long LDPC codes, such structures are known to be responsible of the so-called error floor phenomenon, since their size may be relatively small with respect to the length of the code~\cite{dolecek2009predicting}. However, for short codes, their size may be comparable to the number of errors that the code must correct, therefore possibly inducing a significant degradation of the error correction performance in the waterfall region. To address this issue, we take a decoding diversity approach, implying several BP-RNN decoders working either in serial or in parallel, where each BP-RNN is trained to decode errors corresponding to absorbing sets of a specific type. 
We used a similar approach in our previous work~\cite{rosseel2021error}, where several BP-RNN decoders working in parallel were trained by exhaustively enumerating and classifying (according to the structure of the induced sub-graph) all patterns of two and three errors. The work in~\cite{rosseel2021error} was relevant to high coding rate LDPC codes, correcting a small number of errors. Here, we extend our previous work to lower  rate codes, correcting a higher number of errors, by concentrating our attention to errors with absorbing set support. We further combine our approach with a low-order OSD post-processing step, providing an efficient way to bridge the gap to maximum likelihood (ML) decoding. The main contributions of the paper are summarized below.

\paragraph*{Absorbing set classification and specialization of BP-RNN decoding}  We first propose a graph-search based algorithm, combining backtracking and a deep-first search like procedure, to enumerate, in an efficient way, all the absorbing sets of a given size in the bipartite graph. 
We then perform a fine classification of the enumerated absorbing sets, according to the degree profile of the check-nodes in the induced sub-graph, and train a specific BP-RNN decoder for each absorbing set class. 

\paragraph*{BP-RNNs selection and decoding architectures}   To reduce the number of the BP-RNN decoders, we propose a selection procedure, where the selected decoders are the most complementary in terms of the errors  they can decode. We then consider two decoding architectures, in which the selected BP-RNN decoders are executed in an either parallel or serial manner, and define the appropriate metrics to assess their computational complexity and decoding latency.

\paragraph*{OSD post-processing} 
We further combine our approach with an OSD post-processing step, applied  in case  none of the selected BP-RNN decoders outputs a codeword. There are two motivations behind the use of such a post-processing step. First, for short LDPC codes, the OSD complexity is limited, compatible with practical applications, especially when the OSD order is small. Here, we  restrict the order of the OSD post-processing step to either $0$ or $1$. Second, it may benefit from the diversity brought by the use of multiple BP-RNN decoders. Indeed, we show that the coding gain brought by the use of multiple BP-RNN decoders is actually amplified by the use of the OSD post-processing, resulting in a significant improvement of the error correction performance. 

The paper is organized as follow. Section~\ref{sec:RNN-based-BP} introduces the notation and the BP-RNN decoding algorithm. Absorbing sets, as well as the tree-search based algorithm to enumerate them, their classification, and the training method used to specialize the BP-RNNs to absorbing set classes, are presented in Section~\ref{sec:AS spec}. The BP-RNNs selection, together with the parallel and serial and decoding architectures are presented in Section~\ref{sec:decoding archi}. Section~\ref{sec:OSD} provides the details of the OSD post-processing step. 
Finally, Section~\ref{sec:results} presents the numerical results, and Section~\ref{sec:conclusion} concludes the paper.

\section{Preliminaries}
\label{sec:RNN-based-BP}
\subsection{Neural BP decoding}
\label{ref:BP-RNN}

We consider an LDPC code defined by a Tanner (bipartite) graph~\cite{tanner1981recursive} with $N$ variable-nodes and $M$ check-nodes, denoted respectively by $n \in \{1,\dots,N\}$ and $m \in \{1,\dots,M \}$. 
We further denote by $\mathcal{N}(m)$ the set of variable-nodes connected to a check-node $m$, and by $\mathcal{M}(n)$ the set of check-nodes connected to a variable-node $n$. 

BP decoding consists of an iterative exchange of messages along the edges of the Tanner graph, where each message provides an estimation of the incident variable-node~\cite{savin2014ldpc}. BP-RNN and BP-FF decoding algorithms are weighted variants of the BP decoding. Exchanged messages are multiplied by weights which are learned through an either RNN or FF-NN approach, respectively. The underlying NN contains three types of \emph{neural layers}, each one corresponding to a step of the BP algorithm. The \emph{check-pass layer} and the \emph{data-pass layer} carry out the computation of messages outgoing from check-nodes and variable-nodes, respectively. Each one of them contains a number of neurons equal to the number of edges of the Tanner graph. In addition, the \emph{a~posteriori layer} consists of $N$ neurons, computing the a~posteriori Log Likelihood Ratio (LLR) values of the $N$ variable-nodes. The three layers of the NN are connected such that a check pass layer, followed by a data pass layer and an a~posteriori LLR layer model one iteration of the  BP decoding. In particular, it should be noted that neurons in the check-pass and data-pass layers of the NN correspond to directed edges of the Tanner graph, $m\rightarrow n$ and $n\rightarrow m$, respectively. NN edges between these layers connect two neurons sharing either a common variable-node (\emph{i.e.}, $m\rightarrow n$ and $n\rightarrow m'$, with $m\neq m'$) or a common check-node (\emph{i.e.},  $n\rightarrow m$ and $m\rightarrow n'$, with $n\neq n'$). For more details, we refer to~\cite{nachmani2016learning}.

The formulas below detail the calculation of messages within each layer. We denote by $\beta_{m \rightarrow n}$ and $\alpha_{n \rightarrow m}$  the messages computed by the check-pass and data-pass layers, respectively, and by $\Tilde{L}_{n}$ the messages computed by the a~posteriori LLR layer. The observed (channel) LLR values are denoted by ${L_{\mathrm{ch},n}}$, and are used to initialize $\alpha_{n \rightarrow m}$ messages prior to the first iteration. 
%
\begin{align}
\beta_{m \rightarrow n} &=  2\tanh^{-1}
\left(\prod_{n' \in \mathcal{N}(m)\setminus \{n\}} 
\tanh \left( \frac{\alpha_{n' \rightarrow m}}{2} \right) \right) 
\label{check2var_BP_RNN}\\
\alpha_{n \rightarrow m} &=  L_{\mathrm{ch},n} +  
\sum_{m' \in \mathcal{M}(n)\setminus \{m\}} w_{m' \rightarrow n \rightarrow m} \beta_{m' \rightarrow n} 
\label{var2check_BP_RNN}\\
\Tilde{L}_{n} &=  L_{\mathrm{ch},n} + \sum_{m \in \mathcal{M}(n)} 
\Tilde{w}_{m\rightarrow n} \beta_{m\rightarrow n}
\label{apost_BP_RNN}
\end{align}
It can be observed that weights ($w,\tilde{w}$) are applied only on the NN edges incoming to the data-pass (\ref{var2check_BP_RNN}) and a~posteriori LLR  (\ref{apost_BP_RNN}) layers. Each weight corresponds to one specific edge of the NN. In~(\ref{var2check_BP_RNN}) the weights are denoted by  $w_{m' \rightarrow n \rightarrow m}$, where the subscript indicates both the corresponding neuron $n \rightarrow m$ in the data-pass layer, and the incoming NN edge from neuron $m' \rightarrow n$ in the check-pass layer. In~(\ref{apost_BP_RNN}), the weights are denoted by  $\Tilde{w}_{m\rightarrow n}$, where the subscript indicates  the corresponding neuron $n$ in the a~posteriori LLR layer, and the incoming NN edge from neuron $m \rightarrow n$ in the check-pass layer. For the BP-RNN, the weights only depend on the corresponding edges of the NN, while for the BP-FF they also depend on the iteration number (note that, to simplify notation, we have not indicated the iteration number on the above formulas). 

It is worth noticing that despite a lower number of trainable weights, the BP-RNN achieves similar performance to the BP-FF~\cite{nachmani2018deep}. Therefore, \emph{we will only consider the BP-RNN model}, although our work can be readily adapted to the BP-FF model.  Moreover, to further reduce the number of trainable weights, we use the approach suggested in~\cite{nachmani2018deep}, with the data-pass layer computation modified as,
\begin{equation}
\alpha_{n \rightarrow m} =  L_{\mathrm{ch},n} +  w_{n\rightarrow m}  \sum_{m' \in \mathcal{M}(n)\backslash m} \beta_{m' \rightarrow n}, 
\label{var2check_BP_RNN_BP_friendly_imple}
\end{equation}
where the applied weight $w_{n\rightarrow m}$ only depends on the data-pass neuron $n\rightarrow m$. This simplification reduces the training complexity and makes it possible to reuse conventional BP decoding architectures for an efficient hardware implementation.

\subsection{Loss function}
\label{sec:loss func}

The goal of BP-RNN decoding  is to map the received signal to a valid codeword. This  process can  be interpreted as a classification task, with one class corresponding to one codeword of the code. However, this is practically unrealistic due to the high number of classes~\cite{lian2019learned}. A generally accepted method to avoid this problem is to replace the initial classification task by a simpler binary classification task for each bit. 
To do so, the use of the average binary cross-entropy loss function was suggested in~\cite{nachmani2016learning, nachmani2018deep}.
When both the channel and the decoder are symmetric -- \emph{e.g.}, the additive white Gaussian noise (AWGN) channel and BP-RNN decoder considered in this work -- training (as well as testing) may be performed by assuming the transmitted codeword is the all-zero one. In this case, the average binary cross-entropy loss function simply writes as (see~\cite{nachmani2018deep} for details), 
\begin{align}
\mathrm{Loss}(\Tilde{L}) = -\frac{1}{N}\sum_{n=1}^{N}\log (\sigma(\Tilde{L}_n)),
\label{loss}
\end{align}
where the a posteriori LLR value $\tilde{L}_n$ is taken at the last decoding iteration, and $\sigma(z)=(1 + \exp(-z))^{-1}$ is the sigmoid function, converting the LLR value into the probability that the decoded bit is equal to zero. The loss function~(\ref{loss}) is then minimized during the NN training, thus minimizing the average bit error probability.

Multi-loss functions have also been considered in~\cite{nachmani2016learning, nachmani2018deep} (see also~\cite{lian2019learned}), which take into account the a posteriori LLR values after each decoding iteration. This allows  amplifying the contribution of earlier  iterations to the computed gradient.  In~\cite{nachmani2018deep}, multi-loss optimization has been shown to produce effective gains for high-density BCH codes, with both BP-RNN and BP-FF models. However, when  sparser, cycle reduced parity check matrices where chosen for the same codes, the gain was observed to significantly reduce, or even completely vanish. The same observation was made for check-node regular matrices. In this work we consider short LDPC codes, whose parity-check matrices  are both cycle-reduced and check-node regular. In our experiments, we did not observe any gain brought by multi-loss optimization compared to~(\ref{loss}).  We therefore consider the loss function in~(\ref{loss}) to train our BP-RNN decoders.

\subsection{Number of decoding iterations}
\label{subsec:Number of decoding iterations}

The standard approach for NN-based BP is to both \emph{train} and \emph{test} the decoder for a same maximum number of decoding iterations, usually small, such as $5$ or $10$ iterations (\emph{e.g.},~\cite{nachmani2016learning, nachmani2018deep, lian2019learned}). For the BP-FF model, increasing the number of decoding iterations for training also increases the number of trainable weights, which may become computationally intractable. For the BP-RNN model, this is no longer an issue. Yet, assuming that we want to test the BP-RNN decoder for an increased number of decoding iterations,  a natural question  is whether it is optimal to use the same number of decoding iterations for training. At first sight, this might be the case. However, when the BP decoder does not converge, it may exhibit erratic behavior, especially during the last decoding iterations. Increasing the number of decoding iterations for training may amplify this behavior, and impair the optimization result. We will provide numerical evidence for this in Section~\ref{sec:training ite impact}. For the moment, this motivates us to introduce two parameters, denoted  $I_\text{train}$ and $I_\text{test}$, corresponding to the maximum number of decoding iterations during the training and the testing phases, respectively.

\section{Absorbing Set Classification and Specialization of BP-RNN Decoding}
\label{sec:AS spec}
\subsection{Absorbing sets}
\label{sec:AS}

We consider the definition of absorbing sets from~\cite{dolecek2009predicting,dolecek2010analysis}, given below for the   reader's convenience.

Let $A\subset\{1,\dots,N\}$ be a set of variable-nodes. We consider the set of check-nodes connected to $A$ at least once, which we partition in two disjoint subsets $O(A)$ and $E(A)$, denoting the sets of check-nodes connected to $A$ either an odd or an even number of times, respectively. The set $A$ is said to be an \emph{absorbing set}, if any variable-node in $A$ has strictly more neighbors in $E(A)$ than in $O(A)$. 

According to the above definition, if the variable-nodes in error form as absorbing set, then each of them is connected to a higher number of satisfied check-nodes (not detecting the error) than unsatisfied check-nodes (detecting the error). Such errors are likely to mislead the BP decoder, yielding a decoding failure with high probability.

\begin{figure}
\centering
\subfloat[$P_c = (2,5)$]{\includegraphics[width=.47\linewidth]{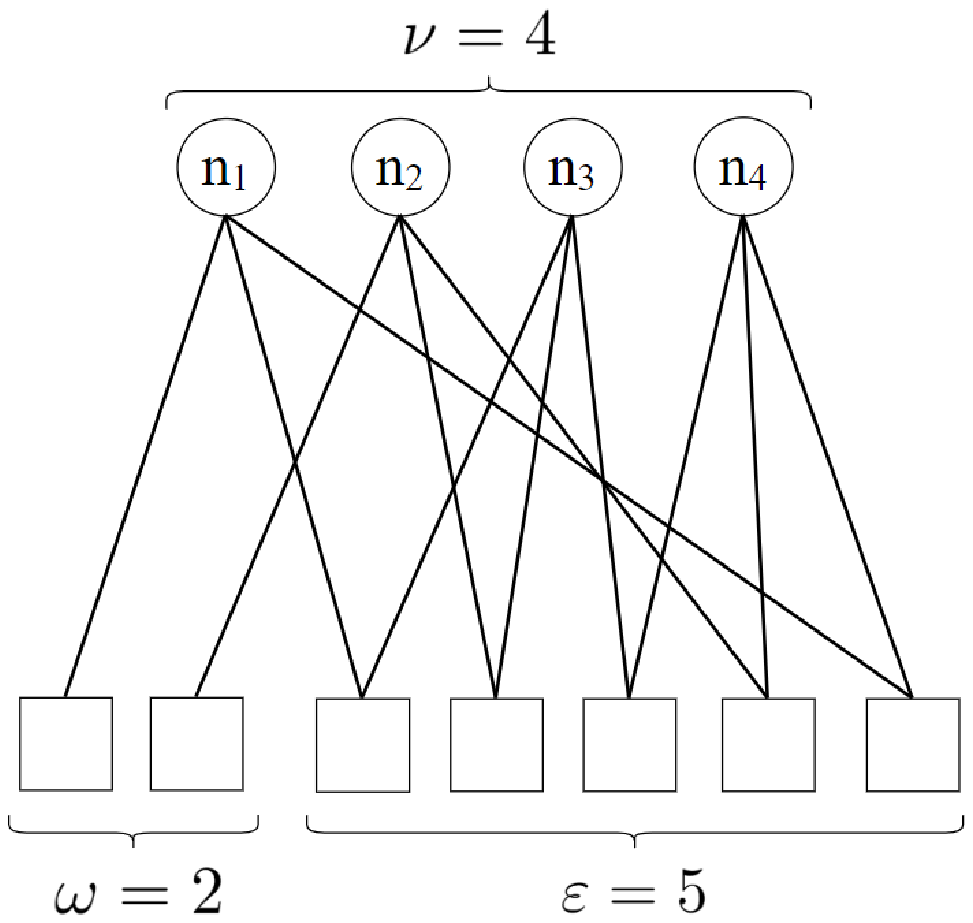}}\quad%
\subfloat[$P_c = (0,5,2)$]{\includegraphics[width=.47\linewidth]{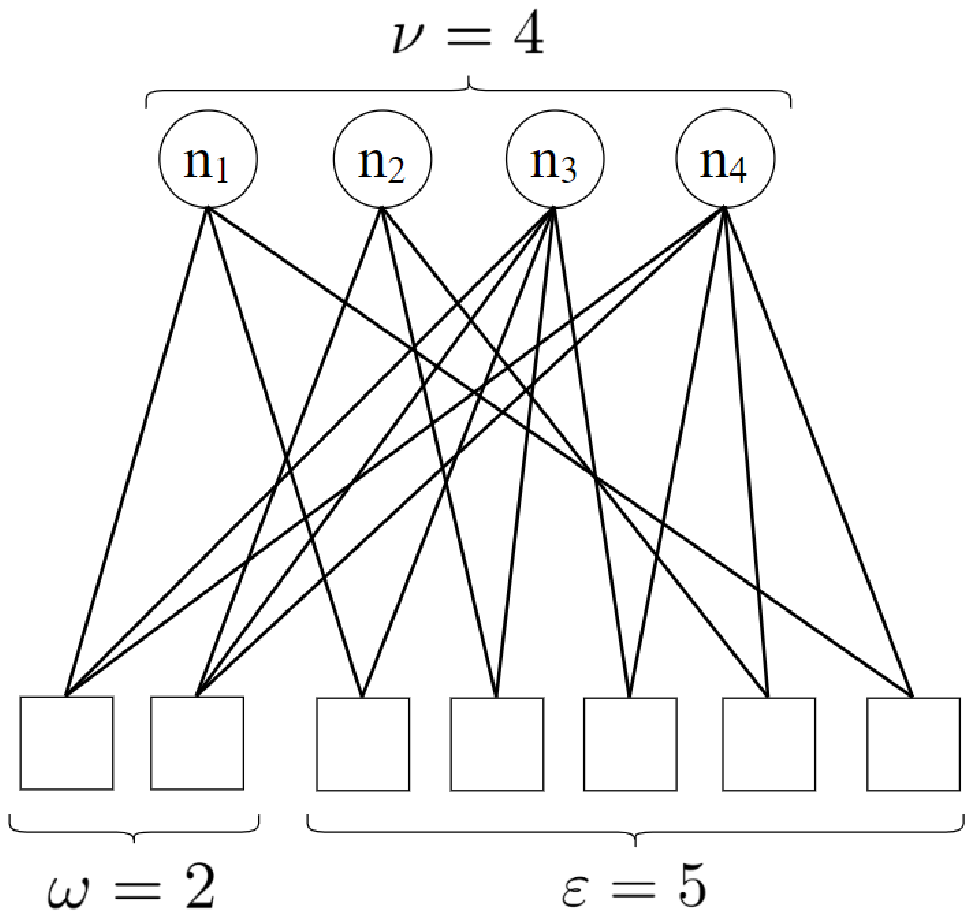}}
\caption{Examples of absorbing sets of type $4\mhyphen(2,5)$}
\label{AS example}
\end{figure}

We say that an absorbing set $A$ is of \emph{type} $\nu\mhyphen(\omega, \varepsilon)$, where $\nu := |A|$, $\omega := |O(D)|$ and $\varepsilon := |E(D)|$, and  $|S|$ denotes the number of elements in the set $S$ (see also Fig.~\ref{AS example}). Thus, the subgraph induced by $A$ comprises $\nu$ variable-nodes and $\omega+\varepsilon$ check-nodes, of which $\omega$ are of odd degree  and $\varepsilon$ are of even degree. Note that absorbing sets of the same type do not necessarily induce the same subgraph. This is illustrated in Fig.~\ref{AS example}, where two absorbing sets of type $4\mhyphen(2,5)$ have different induced subgraphs. 
To further characterize the structure of the induced subgraph, we define the check-node degree profile (of the subgraph induced by $A$) as $P_c = (m_1, m_2, \dots)$, where $m_d$ is the number of check-nodes connected to exactly $d$ variable-nodes in $A$ (note that the sequence $m_d$ is actually finite, of length equal to the maximum check-node degree in the subgraph induced by $A$). Finally, we define the \emph{extended type} of $A$ as $\nu\mhyphen(\omega, \varepsilon, P_c)$.

\subsection{Graph-search based algorithm}
\label{sec:finding AS}

A brute-force search algorithm to enumerate all the trapping/absorbing sets of a given size $\nu$ would have to explore all the $\binom{N}{\nu}$ candidates, which may become computationally intractable even for relatively small values of $N$ and $\nu$. 
Several exhaustive/non-exhaustive enumeration algorithms have been proposed in the literature, to enumerate \emph{elementary} trapping sets (i.e., whose induced subgraphs contain only degree-1 and degree-2 check nodes)~\cite{karimi2014characterization, hashemi2015characterization, falsafain2016exhaustive}, trapping or absorbing sets for specific classes of LDPC codes~\cite{abu2010trapping, dolecek2010analysis} or only small or dominant such structures~\cite{wang2003artificial, karimi2012efficient}, and \emph{fully} absorbing sets (i.e., where the absorbing condition is satisfied by all the variable-nodes in the graph)~\cite{kyung2010exhaustive, zhang2011efficient}. Several of these algorithms rely on a linear programming based branch-and-bound approach, \emph{e.g.},~\cite{falsafain2016exhaustive, kyung2010exhaustive, zhang2011efficient}.

Here, we propose a graph search based algorithm, which, to the best of our knowledge, is the first specifically developed for the exhaustive enumeration of absorbing sets, without any restriction on the structure of the Tanner graph of the absorbing sets to be enumerated.
The proposed algorithm is essentially a backtracking algorithm that incrementally builds absorbing set candidates, and abandons a candidate as soon as it determines that it cannot possibly be completed to an absorbing set. Candidates are built incrementally by traversing the bipartite graph in a depth-first search (DFS) manner, that is, starting from a  variable-node chosen as root, and exploring as deeply as possible before backtracking. The main difference with the standard depth-first search is that our algorithm does not visit only one, but a subset of variable-nodes at each depth level, which are chosen to increment the candidate solution.  

Let $n\in\{1,\dots,N\}$ be a fixed variable-node, which we will refer to as root node. Our algorithm enumerates all the absorbing sets $A\subset\{1,\dots,N\}$ containing $n$, of size $|A| = \nu$. To avoid enumerating a same absorbing set multiple times (\emph{i.e.}, starting with different root nodes), we will further require $n$ to be the smallest element of $A$, that is, $A\subset\{n,\dots,N\}$.

We first expand the bipartite graph starting from the root node $n$, and incrementally adding check and variable nodes at an increasing distance\footnote{The distance between two nodes is the length (\emph{i.e.}, number of edges) of a shortest path connecting them.}  from $n$, until the  most distant node in the bipartite graph is reached. This expansion produces a \emph{rooted bipartite graph}, which we will denote by $\mathcal{H}$.
For $\ell\geq0$, we denote by $\mathcal{V}_\ell$ the set of variable-nodes at distance $2\ell$ from $n$ (thus, $\mathcal{V}_0 = \{n\}$), and by  $\mathcal{C}_\ell$ the set of check-nodes at distance $2\ell+1$ from $n$.   
%
Let $A$ be any set of variable-nodes, containing $n$, and let $C$ be the set of check-nodes connected to $A$ at least once. For $\ell \geq 0$, we define $A_{\ell} := A \cap \mathcal{V}_{\ell}$ and $C_{\ell} := C \cap \mathcal{C}_{\ell}$. We say that  $A_\ell$ satisfies the absorbing set condition, and we write $\mathop{\rm AS\_check} (A_{\ell}) = \text{true}$, if  any variable-node in $A_\ell$ has strictly more neighbors in $E(A)$ than in $O(A)$. Now, the neighbors of the variable-nodes in $A_\ell$ are check-nodes that belong to either $C_{\ell-1}$ or $C_{\ell}$, and whose degrees, in the subgraph induced by $A$, only depend  on  $A_{\ell-1}$ (for $\ell > 0$), $A_\ell$,  and $A_{\ell+1}$ (for $\ell$ less than the maximum depth of $A$). Hence, one may check the absorbing set condition for $A$ in an incremental manner, by checking it for each subset $A_\ell$, which however requires the knowledge of the next level subset $A_{\ell+1}$ (except for the last level).

The proposed algorithm incrementally builds an absorbing set $A$,  by choosing subsets $A_{\ell}$, for $\ell \geq 0$, and checking the absorbing set condition to determine whether  the current choice could possibly be completed to a valid absorbing set. For $\ell = 0$, we have $A_0 = \{n\}$. Consider some choice of subsets $A_0, \dots, A_{\ell}$, with $\ell \geq 0$, such that $\bar{\nu}_\ell := \sum_{t=0}^\ell |A_t| \leq \nu$. We define the \emph{set of possible completions} $\mathfrak{C}(A_0, \dots, A_{\ell})$, as follows:
\begin{itemize}
\item if $\bar{\nu}_\ell = \nu$, then $\mathfrak{C}(A_0, \dots, A_{\ell}) = \emptyset$ (absorbing set size $\nu$ already reached, thus, no completion needed).
\item if $\bar{\nu}_\ell < \nu$, then \\
(1) determine the subset $\mathcal{N}_{\ell+1}\subset  \mathcal{V}_{\ell+1}$ containing  the descendants of the variable-nodes in $A_{\ell}$. \\
(2) set $\mathcal{N}_{\ell+1}^{(\geq n)} := \mathcal{N}_{\ell+1} \cap \{n,\dots,N\}$, and
\end{itemize}
\begin{equation}
\mathfrak{C}(A_0, \dots, A_{\ell}) := \left\{ A_{\ell+1} \subseteq \mathcal{N}_{\ell+1}^{(\geq n)} \,\middle|\, |A_{\ell+1}| \leq \nu - \bar{\nu}_\ell \right\}.
\end{equation}
Note that $\mathfrak{C}(A_0, \dots, A_{\ell})$ is a set of subsets $A_{\ell+1}$ (which may be empty, if either $\bar{\nu}_\ell = \nu$ or $\mathcal{N}_{\ell+1}^{(\geq n)} = \emptyset$), and which we may iterate through in any convenient order, \emph{e.g.}, according to increasing  size of subsets $A_{\ell+1}$.  

The procedure~\ref{algo:AS-Enumeration} $\mathop{\rm AS\_DFS}$ provides a recursive implementation of the proposed algorithm. Given a variable-node $n$, the procedure is simply called from the main program with inputs $\mathop{\rm AS\_DFS}(\mathcal{H}, \ell=0, A_0 = \{n\})$.

\begin{algorithm}[!t]
\caption{$\mathop{\rm AS\_DFS}(\mathcal{H}, \ell, A_0,\dots, A_\ell)$}
\label{algo:AS-Enumeration}
\begin{algorithmic} 
\If{$\ell > 0$ and $\mathop{\rm AS\_check} (A_{\ell-1}) = \text{false}$}
	\State \Return \ // backtrack
\EndIf

\smallskip
\State $\bar{\nu}_\ell \gets |A_0| + \cdots + |A_\ell|$

\If{$\bar{\nu}_\ell = \nu$}
	\If{$\mathop{\rm AS\_check} (A_{\ell}) = \text{true}$}
		\State Add $A := A_0\cup \cdots \cup A_{\ell}$ to the list of absorbing sets
	\EndIf
	\State \Return \ // backtrack
\EndIf

\smallskip
\State $\mathfrak{C}(A_0, \dots, A_{\ell}) \gets$ set of possible completions
\For{\textbf{all} $A_{\ell+1}$ \textbf{in}  $\mathfrak{C}(A_0, \dots, A_{\ell})$}
	\State $\mathop{\rm AS\_DFS}(\mathcal{H}, \ell+1, A_0,\dots, A_\ell, A_{\ell+1})$ \ // recursive call
\EndFor
\end{algorithmic}
\end{algorithm}

To illustrate the capability of the proposed algorithm, we consider the following two codes, both of rate $1/2$, which will be subsequently used throughout this paper.

\begin{description}
\item[\textbf{Code-1}] is a regular LDPC code, of length $64$ bits, with variable-nodes of degree $3$ and check-nodes of degree $6$, constructed by using the progressive edge growth (PEG) algorithm~\cite{hu2005regular}. It has girth $g=6$, with multiplicity $164$ (\emph{i.e.}, number of cycles of length $g$). 
\item[\textbf{Code-2}] is the  LDPC code from~\cite{codeCCSDS2015}, known as the CCSDS LDPC code. It is of length $128$ bits, with half of variable-nodes of degree $3$ and the other half of degree $5$, and check-nodes of degree $8$. It has  girth $g=6$, with multiplicity $2336$.
\end{description}

In Table~\ref{tab:as-enumeration}, we provide the total number of different extended-type (ET) values and the total number of absorbing sets, for both Code-1 and Code-2, and absorbing set size values $\nu \leq 8$. We note that for Code-2 and $\nu=8$, a brute-force search algorithm would require exploring a number of $\binom{128}{8} \approx 2^{40}$ candidates. Our algorithm enumerated all the absorbing sets of size $\nu=8$ in $38$ minutes (Intel Xeon @2.20GHz processor).
 
 \begin{table}
 \caption{Number of Extended-Types (ET) and Absorbing Sets (AS)}
 \label{tab:as-enumeration}
 \centering
{\small\begin{tabular}{|c|@{\ }c@{\ }|@{\ }c@{\ }|@{\ }c@{\ }|@{\ }c@{\ }|}
 \hline
 & \multicolumn{2}{c@{\ }|@{\ }}{Code-1} & \multicolumn{2}{@{\ }c@{\ }|}{Code-2} \\
 \hline
 $\nu$ &  ET-Number & AS-Number & ET-Number & AS-Number \\
 \hline\hline
   3   &      1     &    164    &     1     &       32  \\
   4   &      2     &   1\,452    &     6     &      944  \\
   5   &      3     &   9\,413    &    12     &    11\,504  \\
   6   &      9     &   64\,813$^*$   &    32     &   152\,824  \\
   7   &     16     &   450\,340  &    69     &  2\,124\,928  \\
   8   &     24     &  2\,994\,834$^*$  &     157      &  28\,670\,736 \\
   \hline
   \multicolumn{5}{@{\,}l@{\,}}{\vbox{\vspace*{1mm}
  \hbox{\footnotesize $^*$Code-1 contains one codeword of weight $6$, and $37$ codewords}
  \hbox{\footnotesize \hspace*{2mm}of weight $8$, corresponding to absorbing sets with $\omega=0$}}}
\end{tabular}}
\end{table}

\subsection{Training specialized BP-RNN decoders}
\label{sec:Training set}
Having at our disposal an efficient algorithm to enumerate absorbing sets of a given size, we perform a fine classification of the found absorbing sets, by grouping into a same class absorbing sets with the same extended type. The goal is to train a specific BR-RNN decoder for each class. The approach bears similarity to, and is  motivated by~\cite{declercq2013finite, xiao2021faid}, where decoding rules for FAIDs have been either designed or learned to correct specific trapping sets for the binary symmetric channel. 

We consider a binary-input AWGN channel, with BPSK alphabet ($\pm 1$) inputs, and fixed noise variance $\sigma^2$. Assuming the all-zero codeword is transmitted (see Section~\ref{sec:loss func}), corresponding to the $+1$ input of the channel, the received word is given by 
\begin{align}
\label{AWGN model}
y_n = 1 + z_n, \quad z_n \sim \mathcal{N}(0, \sigma^2),\ n=1,\dots,N,
\end{align}
where $\mathcal{N}(0, \sigma^2)$ denotes a normal distribution, with mean $0$ and variance $\sigma^2$. The signal-to-noise ratio (\snr) is defined as $\snr = -10\log_{10}(\sigma^2)$\,dB.

Given a received word $y:=(y_1,\dots, y_N)$, we define its \emph{error set} as $E(y) := \{n \mid y_n \leq 0\} \subset \{1,\dots,N\}$. Our goal is to construct a specific training set, containing only words $y$, such that $E(y)$ is an absorbing set of a given extended type $\nu\mhyphen(\omega, \varepsilon, P_c)$. 
As a neural network specializes to \emph{trends} embedded in the training set~\cite{wang2003artificial}, we expect the trained BP-RNN to specialize in decoding words $y$, such that $E(y)$ is, or possibly contains, an absorbing set of extended type $\nu\mhyphen(\omega, \varepsilon, P_c)$.

One way to generate such a training set is to randomly generate words $y$ according to~(\ref{AWGN model}), and to select those for which $E(y)$ is an absorbing set of extended type $\nu\mhyphen(\omega, \varepsilon, P_c)$. This would be rather tedious and time consuming. A more efficient way is to use our knowledge of absorbing sets (determined by the algorithm from Section~\ref{sec:finding AS}), and generate Gaussian noise by means of a truncated Gaussian distribution, to produce errors only on desired locations.  Precisely, we proceed as follows. We first chose an absorbing set $A$ at random, from those having the desired extended type $\nu\mhyphen(\omega, \varepsilon, P_c)$. Then we generate a random  word $y = (y_n = 1+z_n)_{n=1,\dots,N}$, with 
\begin{align}
\label{training set construction}
z_{n} &\sim \left\{
    \begin{array}{ll}
        \mathcal{N}(0, \sigma^2, -\infty, -1), & \text{if }   n \in A \\
        \mathcal{N}(0, \sigma^2, -1, \infty), & \text{otherwise}
    \end{array} 
\right. 
\end{align}
where $\mathcal{N}(0, \sigma^2, a, b)$ denotes the truncated normal distribution with mean $0$ and variance $\sigma^2$, taking values in the interval $(a,b)$. Clearly, we have $E(y) = A$. The training set is obtained by repeating the above procedure (including the random choice of $A$)  multiple times. This way, the training set is representative of the \emph{error class} (identified by the  extended type of the absorbing error set), and thus the trained BP-RNN decoder becomes specialized in decoding the errors of the class.

\begin{figure}[!t]
\centering
\subfloat[Data pass layer ]{\includegraphics[height=4.5cm]  {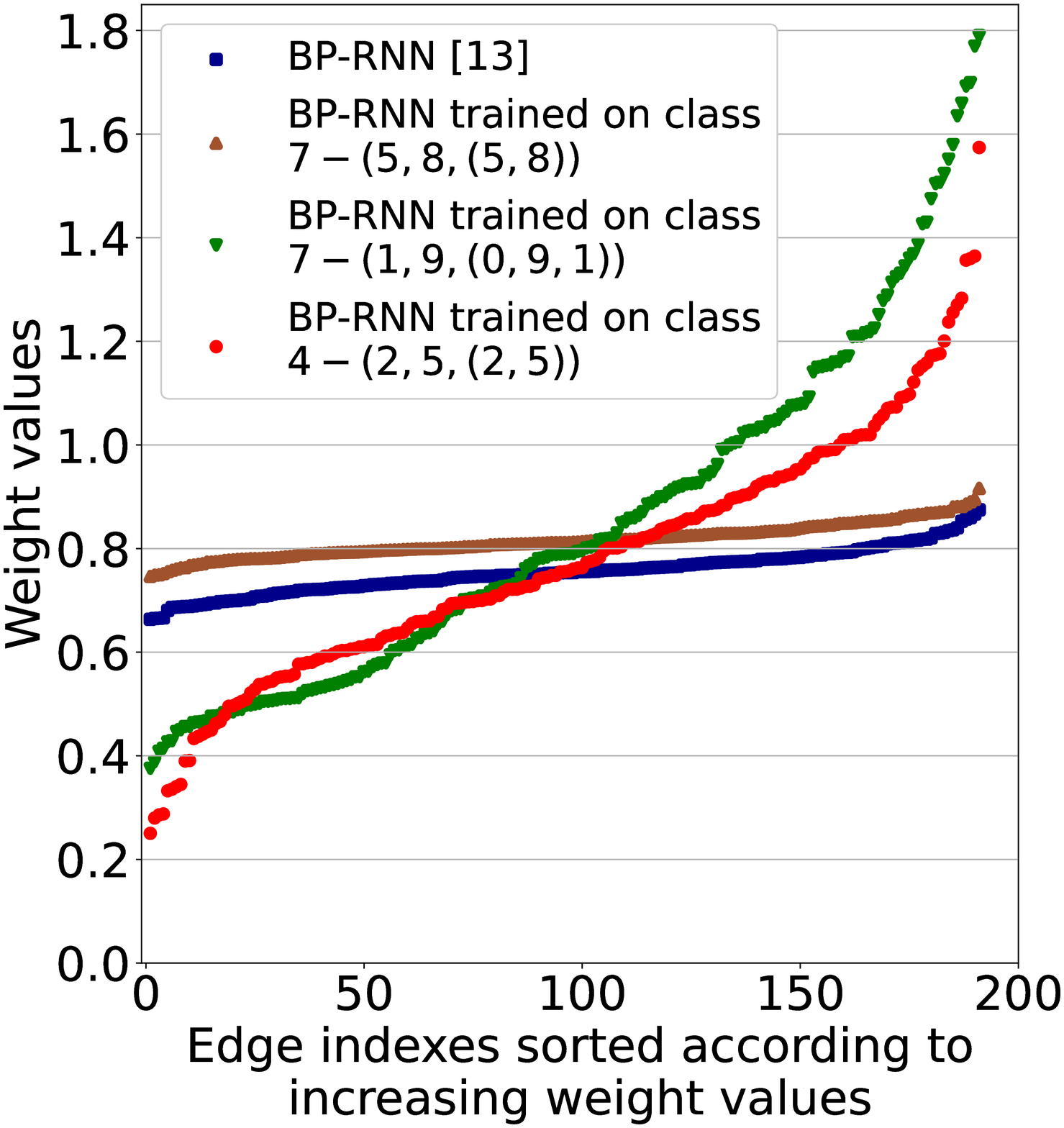}%
  \label{Weight profil data pass later}}
    \quad%
    \subfloat[A posteriori layer ]{\includegraphics[height=4.5cm]{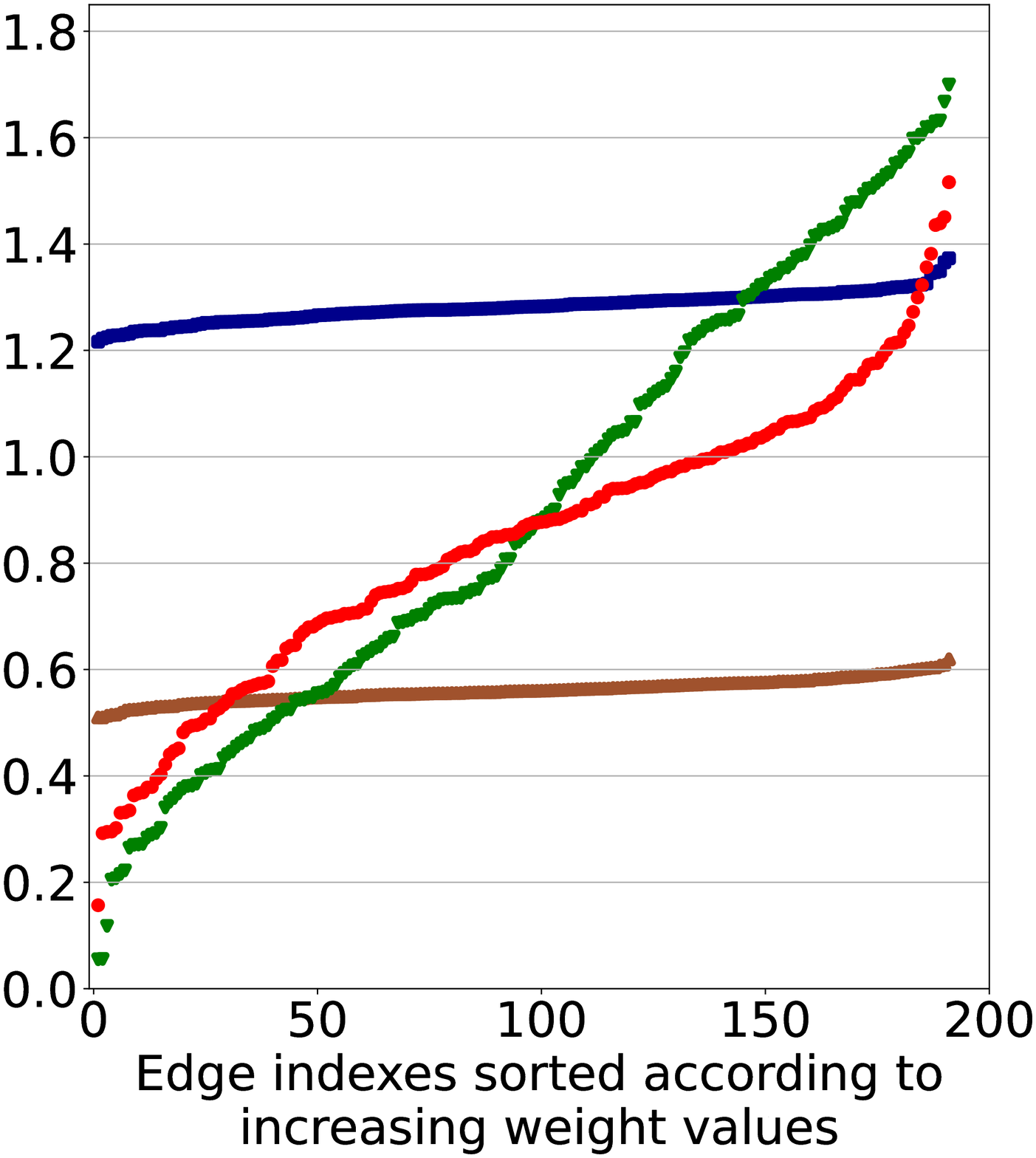} %
    \label{Weight profil apost pass later}}
    \caption{Weight profiles of various BP-RNN decoders (Code-$1$, $\snr = 5$\,dB)}
    \label{Weight profiles comple}
\end{figure}

To illustrate the specialization of the trained decoder, Fig.~\ref{Weight profiles comple}  shows the weight values for three BP-RNN decoders, trained for three different error classes, with corresponding extended types shown in the legend. The (unspecialized) BP-RNN decoder trained as in~\cite{nachmani2018deep} is also shown.  To avoid clutter, weight values are sorted in increasing order, and an ordered set of weight values is referred to as  \emph{weight profile}\footnote{Note that different weight profiles indicate different sets of weight values, while similar weight profiles indicate similar sets of weight values. However, in the latter case, similar weights may apply to different edges.}. We consider the Code-1 and show the weight profiles for the both data pass and a posteriori layers. The data pass layer is implemented as in~(\ref{var2check_BP_RNN_BP_friendly_imple}), with a same weight applied on all the incoming edges to each neuron (see Section~\ref{sec:results} for a complete list of training parameters). Hence, the number of weights is the same for  both the data pass and the a posteriori layers, equal to the number of edges of the Tanner graph ($192$ for Code-1). The weight profiles in Fig.~\ref{Weight profiles comple} clearly indicate that  weight optimization responds differently to training sets corresponding to different error classes. We exploit in the next section the diversity induced by the BP-RNN specialization.

\section{BP-RNNs Selection and Decoding Architectures}
\label{sec:decoding archi}

\subsection{BP-RNN diversity selection}
\label{subsec:BP-RNN diversity selection}
Using the procedure from Section~\ref{sec:Training set}, we train one specialized BP-RNN decoder for  each error class, \emph{i.e.}, extended type\footnote{We take off extended types for which $\omega=0$, as they correspond to the support of non-zero codewords. Such errors  cannot be detected or corrected.} $\nu\mhyphen(\omega, \varepsilon, P_c)$, with absorbing error set size $\nu \leq \nu_\text{max}$ (the choice of $\nu_\text{max}$ is discussed in Section~\ref{sec:results}). Let $J$ denote the total number of specialized BP-RNN decoders.

Here, we propose a selection procedure to reduce the number of BP-RNN decoders, as well as to avoid similar training effects (which may arise, for instance,  due to absorbing sets of a given size containing absorbing sets of smaller sizes). To do so, we assess the complementarity of the trained decoders, in terms of the errors  they can decode. 

We first generate a common test set, $T$, containing random  words $y$ defined as in~(\ref{AWGN model}), which is then used to assess the individual error correction performance of each of the trained decoders.  We denote by $F_j \subset T$ the susbset of words on which the BP-RNN decoder $D_j$ failed, where $j=1,\dots, J$. 
Then, we recursively construct an \emph{ordered list} of decoder indexes, denoted $\mathfrak{J}$, as follows.  We start by initializing  $\mathfrak{J}$ as the empty list,  $\mathfrak{J} = \emptyset$. To add a new index, $j_\text{new}$ to $\mathfrak{J}$, we use the following rule,
\begin{align}
\label{select rule}
j_\text{new} = \argmin_{j \in \{1,\dots,J\}\setminus \mathfrak{J}} \left| F_{\mathfrak{J}} \cap F_j \right|,
\end{align}
where $F_{\mathfrak{J}} := T$, if $\mathfrak{J} = \emptyset$, and $F_{\mathfrak{J}} := \cap_{i\in \mathfrak{J}} F_i$, otherwise. The above rule is applied recursively $J$ times, until $\mathfrak{J}$ contains all the decoder indexes $j=1,\dots,J$, in a sorted order. In case the argmin in~(\ref{select rule}) is not unique, an arbitrary one is chosen.

Note that when $\mathfrak{J} = \emptyset$,  the rule (\ref{select rule}) rewrites as $j_\text{new} = \argmin_{j \in \{1,\dots,J\}} \left| F_j \right|$. Hence, the first decoder (index) added to the list is the one minimizing the word error rate. Subsequently,  when $\mathfrak{J} \neq \emptyset$, the new decoder added to the list is the most \emph{complementary} with those already in $\mathfrak{J}$, in the sens that it minimizes the number of words on which all the decoders indexed by $\mathfrak{J} \cup \{j_\text{new}\}$ fail.

For $Z \leq J$, let $\mathfrak{J}(1\mathord{:}Z) \subset \mathfrak{J}$ be the sublist defined by the first $Z$ indexes in $\mathfrak{J}$. We define
\begin{align}
\label{eq:diversity_size_Z}
\mathcal{D}_Z := \left\{  D_j \mid j \in \mathfrak{J}(1\mathord{:}Z) \right\}
\end{align}
Note that $\mathcal{D}_Z$ is an \emph{ordered list of decoders}\footnote{As a matter of fact, the three  BP-RNN decoders in Fig.~\ref{Weight profiles comple}, specialized on absorbing set error classes, correspond to  $\mathcal{D}_{Z=3}$.}, which we will  refer to as \emph{BP-RNN diversity} of size $Z$ (using a similar terminology  to the one in~\cite{declercq2013finite, xiao2021faid}). The $Z$ BP-RNN decoders in  $\mathcal{D}_Z$ may  then be used with  either a parallel or a serial decoding architecture, as discussed in the next section. 
The value of $Z$ may be  dictated by complexity reasons, or, as illustrated in Section~\ref{sec:comple results}, chosen to ensure small (negligible) degradation of the error correction performance, with respect to the case when all $J$ BP-RNN decoders are used.

\subsection{BP-RNN diversity decoding architectures}
\label{subsec:bprnn_diversity_architectures}

\begin{figure*}
\centering
\subfloat[Parallel architecture]{\includegraphics[scale=0.57]{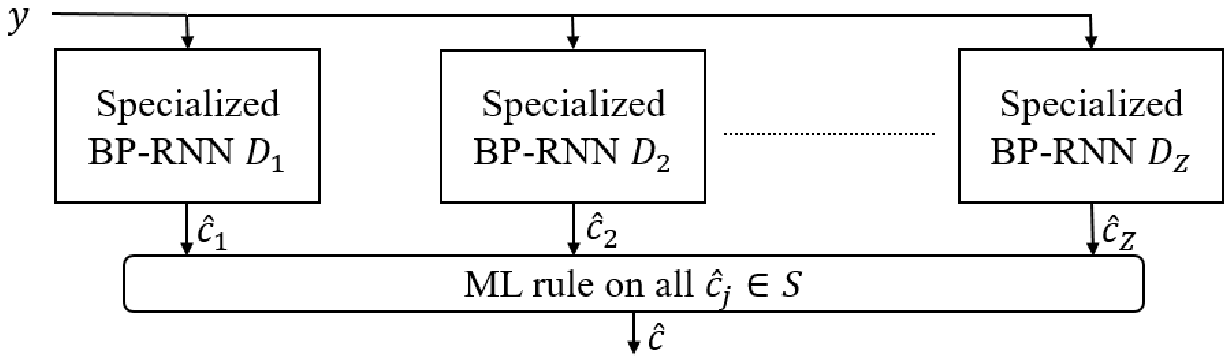}\label{fig:bprnn-diversity-parallel}}\hfill%
\subfloat[Serial architecture]{\includegraphics[scale=0.57]{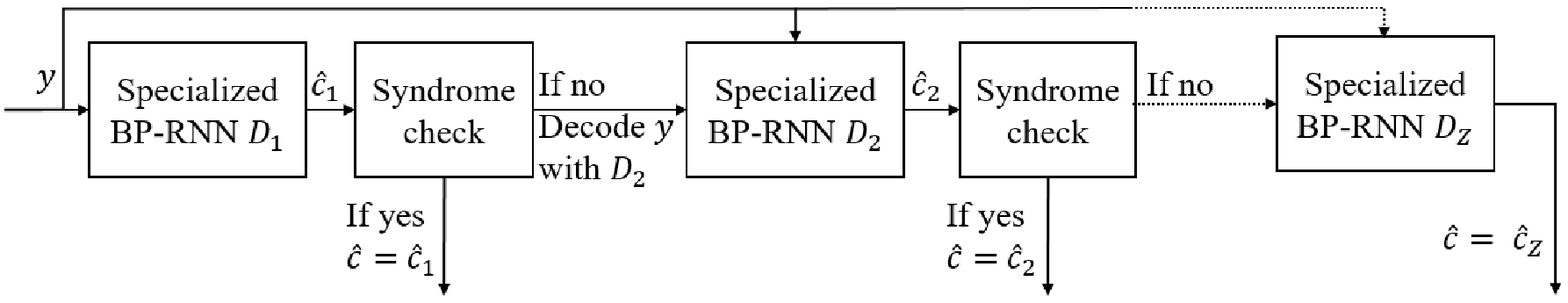}\label{fig:bprnn-diversity-serial}}
\caption{BP-RNN diversity decoding architectures}
\label{fig:bprnn-diversity-architectures}
\end{figure*}

We consider a BP-RNN diversity $\mathcal{D}_Z$, comprising $Z$ BP-RNN decoders. For simplicity,  we  index the BP-RNN decoders in $\mathcal{D}_Z$ from $1$ to $Z$, thus $\mathcal{D}_Z = \{D_1,\dots,D_Z\}$. Fig.~\ref{fig:bprnn-diversity-architectures} shows the proposed parallel and serial architectures, using the $Z$ BP-RNN decoders in $\mathcal{D}_Z$.

In the parallel architecture, each decoder outputs an estimate $\hat{c}_j = (c_{j,1},\dots,c_{j,N})\in\{0,1\}^N$ of the transmitted codeword $c$, according to the sign of the corresponding LLR values at the last decoding iteration. The output of the parallel architecture  is determined as the maximum likelihood (ML) codeword, among the codewords outputted by the constituent BP-RNN decoders (if any, see below).  For the binary-input AWGN channel (with $\pm 1$ inputs), the ML criterion simply writes as
\begin{align}
\hat{c}= \argmax_{\hat{c}_j \in S} P(\hat{c}_j|y) 
       = \argmin_{\hat{c}_j \in S} \sum_{n=1}^N  y_n \hat{c}_{j,n},
\label{ML rule}
\end{align}
where $S := \{\hat{c}_j \mid \synd(\hat{c}_j) = 0\}$ denotes the set of $\hat{c}_j$'s verifying the syndrome. Decoding is then successful if $\hat{c}$ is equal to the transmitted codeword. If none of the BP-RNN decoders outputs a codeword, the decoding  fails. In such a case, the ML criterion -- or a similar bitwise  maximum a posteriori criterion -- may still be applied to select one of the  $\hat{c}_j$ outputs, if desired, \emph{e.g.}, in order to minimize the bit error rate of the decoder  (however, we will only be concerned with word error rate results in this paper).

In the serial architecture, the constituent BP-RNN decoders are run sequentially according to the order given by the sorting procedure from Section~\ref{subsec:BP-RNN diversity selection}. Decoding stops as soon as a BP-RNN decoder $D_j$ outputs a codeword $\hat{c}_j$ ($\synd(\hat{c}_j) = 0$), which becomes the output $\hat{c}$ of the serial architecture.  Decoding is then successful if $\hat{c}$ is equal to the transmitted codeword. If none of the BP-RNN decoders outputs a codeword,  decoding  fails. For simplicity, in Fig.~\ref{fig:bprnn-diversity-serial}, we take $\hat{c}_Z$ as the output of the serial architecture in this case.

The parallel architecture yields a reduced maximum decoding latency, compared to the serial one. This comes however at the cost of an increased complexity, from the both  computational and hardware  perspectives, since  $Z$ BP-RNN decoders must be instantiated in hardware. For the serial architecture, only one decoder may be instantiated in hardware, which may  then be reused to perform sequentially the $Z$ BP-RNNs (while updating the corresponding set of weights). The average-case computational complexity, as well as the average-case decoding latency, for the both parallel and serial architectures are discussed in the next section.

\subsection{Average-case complexity and decoding latency metrics}
\label{sec:comple decoding latency}

We consider a BP-RNN diversity  $\mathcal{D}_Z$, where all the constituent decoders are set with a maximum number of decoding iterations $I_\text{test}$ (see Section~\ref{subsec:Number of decoding iterations}). For a BP-RNN decoder $D_j$, we denote by $I_{{D_j},y} \leq I_\text{test}$  the number of iterations performed by $D_j$ to decode $y$, and  define $\overline{I}_{D_j}:=\mathbb{E}_y(I_{{D_j},y})$, where $\mathbb{E}$ denotes the expected value operator. In practice, $\overline{I}_{D_j}$ is estimated by averaging over the decoded words $y$. We further define the average number of decoding iterations of a decoding (parallel or serial) architecture using $\mathcal{D}_Z$, as  
\begin{align}
\overline{I}({\mathcal{D}_Z}) = \sum_{D_j \in \mathcal{D}_Z} \overline{I}_{D_{j}}.
\label{eq:ave_ite_para}
\end{align}
Note that for the serial architecture, $I_{{D_j},y}$ may be equal to zero, in case the decoding process stops before reaching ${D_j}$. Thus, $\overline{I}_{D_j}$ may be close to $0$, if $D_j$ is rarely used (unlike the parallel architecture, which uses all decoders). Since the computational complexity scales linearly with the number of decoding iterations  performed by the decoder, we use the average number of decoding iterations $\overline{I}_{\mathcal{D}_Z}$ as a measure of the \emph{average-case computational complexity} of the BP-RNN diversity architecture.

For the serial architecture, $\overline{I}_{D_j}$ may also be seen as a measure of the \emph{average-case decoding latency}, since decoding iterations are performed sequentially within each constituent decoder, while the constituent decoders  are also run sequentially. However, this is no longer the case for the parallel architecture. Hence, we define the decoding latency of the parallel architecture, for decoding a given word $y$, as
\begin{align}
L_{\mathcal{D}_Z, y} = \max_{D_j \in \mathcal{D}_Z} I_{D_{j},y},
\label{decoding latency_single_word_para}
\end{align}
corresponding to the maximum number of iterations performed by the constituent decoders to decode $y$. Finally, the average-case decoding latency of the parallel architecture is defined as 
\begin{align}
\bar{L}({\mathcal{D}_Z}) := \mathbb{E}_y(L_{\mathcal{D}_Z, y}),
\label{eq:ave_lat}
\end{align}
and is estimated by averaging over the decoded words $y$.

\section{OSD Post-Processing for BP-RNN Decoders}
\label{sec:OSD}

OSD was first proposed in~\cite{fossorier1995osd}, as a decoding method capable to approach the ML decoding performance, for moderate-length linear block codes, with polynomial complexity. It can be used as a stand-alone decoding algorithm, exploiting the soft-output of the channel ($L_{\mathrm{ch},n}$), or as a post-processing step, exploiting the output of a soft-decision decoder ($\tilde{L}_n$, see Section~\ref{ref:BP-RNN}, for the notation). 

In OSD, variable-nodes are first sorted according to their
\emph{reliability} (that is the absolute value of the corresponding soft decision). The parity-check matrix of the code is then brought to a systematic form\footnote{For simplicity, we assume here that the parity check-matrix is of rank $M$.}, $H = [A \mid I]$, where $A$ is a matrix of size $M \times (N-M)$ and $I$ is the identity matrix of size $M \times M$, and so that the $K:=N-M$ columns of $A$ correspond to the most  possible\footnote{Taking into account that column swaps may be needed, in case the $M$ least reliable columns of $H$ are not linear independent.}  reliable variable-nodes. 
By a slight abuse of language, we simply refer to variable-nodes corresponding to the columns of $A$ as the most reliable ones, and to the remaining variable-nodes as the least-reliable ones. In OSD-0, hard-decision is made on the most reliable variable-nodes, and the least reliable ones are determined by solving the linear system given by $H$.  Hence, decoding is successful if and only if the most reliable variable-nodes are error-free. To address the case where these variable-nodes contain  errors, OSD-$w$ considers all the possible choices of at most $w$ errors among them. For each choice, the initial hard-decision of the corresponding variable-nodes is flipped, and the least reliable variable-nodes are determined again by solving the linear system given by $H$. This procedure produces a list of $\sum_{i=0}^{w} \binom{K}{i}$ codewords, from which the most likely one is selected, according to an ML rule, such as~(\ref{ML rule}). OSD-$w$ may closely approach the  ML decoding performance, assuming the $w$ value (referred to as OSD order) is suitably large.

To bridge the error correction performance gap between suboptimal BP decoding and ML decoding, \cite{fossorier2001iterative} suggested combining  BP decoding with a low-order OSD ($w\leq 1$), where an OSD step is performed at the end of each iteration of the BP decoding.  Here, we propose the use of OSD as a \emph{post-processing step}, applied only in case that none of the constituent BP-RNN decoders (of the BP-RNN diversity $\mathcal{D}_Z$) outputs a codeword. In such a case, we process one OSD using the soft-decision  (a posteriori LLRs) delivered by each  of the constituent BP-RNN decoders. This produces a list of $Z$ codewords (one for the OSD post-processing of each BP-RNN decoder), and the ML rule~(\ref{ML rule}) is used to determine the most-likely one, which becomes the outputted codeword $\hat{c}$. Note that the above description applies to the both parallel and serial architectures, since  OSD post-processing is only performed when all the constituent BP-RNN decoders failed to find a codeword. To reduce the complexity of the post-processing step, we also limit the order of the OSD to $w\leq 1$. 

\begin{figure}[!t]
    \includegraphics[width=.7\linewidth]{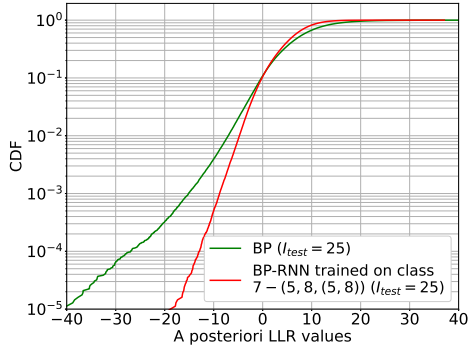}
    \centering
    \caption{CDF of the a posteriori LLR values when decoding fails (Code-1, $\snr = 4$\,dB, maximum number of decoding iterations $I_\text{test} = 25$)}
    \label{cdf code 1}
\end{figure}

There are two main motivations behind the use of the OSD post-processing step. First, the complexity of the low-order OSD is dominated by the Gaussian elimination step, needed to bring the parity check-matrix to a systematic form. However, for LDPC codes, the sparsity of the parity-check matrix can be advantageously exploited to significantly reduce the complexity of this step. See for instance the method proposed in~\cite{lamacchia1991solving} for solving sparse linear systems, which has  been adapted in~\cite{richardson2001efficient} to derive an efficient encoding technique for LDPC codes, and  in~\cite{burshtein2004efficient} to derive an efficient ML decoding algorithm for LDPC codes over erasure channels (the situation is similar for OSD, where the least reliable variable-nodes can be seen as being erased). Second, we use OSD to post-process the soft-output of BP-RNN decoders. Since the loss function~(\ref{loss}) used to train these decoders penalizes negative LLR values, corresponding to variable-nodes in error\footnote{Recall the all-zero codeword assumption from Section~\ref{sec:loss func}.}, we expect such negative LLR values to have a reduced amplitude, thus reducing the probability of error on the most reliable variable-nodes. This is illustrated in Fig.~\ref{cdf code 1}, where we plot the cumulative distribution function (CDF) of the a-posterori LLR values for the BP decoder, and for the BP-RNN decoder trained on the error class $7\mhyphen(5,8,(5,8))$. Moreover, we also expect OSD post-processing to benefit from the diversity brought by the use of multiple BP-RNN decoders, increasing the probability that at least one of these decoders has at most $w$ errors among the most-reliable variable-nodes.

\section{Numerical results}
\label{sec:results}

\subsection{Training settings}
\label{sec:settings}

We consider the two LDPC codes of rate $1/2$ and length either $64$ (Code-1) or $128$ (Code-2) bits, detailed in Section~\ref{sec:finding AS}. We train one specialized BP-RNN decoder for  each absorbing set error class, with error set size $\nu \leq \nu_\text{max}$. 
\begin{itemize}
\item For Code-1, we choose $\nu_\text{max} = 8$, which gives a total number of $J=52$ error classes (see Table~\ref{tab:as-enumeration}\,\footnote{Note that the total number of error classes (ET-Number) in~Table~\ref{tab:as-enumeration}  is $55$, three of which correspond to the support of non-zero codewords of size either $\nu=6$ or $\nu=8$, for which training is not performed.}).
\item For Code-2, we choose $\nu_\text{max} = 7$, which gives a total number of $J=120$ error classes (see Table~\ref{tab:as-enumeration}).
\end{itemize}

For Code-2, the choice of $\nu_\text{max} = 7$ is due to complexity reasons (to limit the number of trained decoders). However, we note that for an $\snr=4$\,dB, in the waterfall region of Code-2, the average number of errors is  $Np_e = 7.2$, where $N=128$ is the code-length, $p_e = Q(1/\sigma) =0.0565$ is the error probability of the binary-input AWGN channel, $\sigma$ is the standard deviation of the Gaussian noise, and $Q$ denotes the Q-function. For Code-1, for the same $\snr=4$\,dB, the size of a random error set is less than or equal to the chosen $\nu_\text{max}=8$, with probability slightly greater than $0.99$.

All the numerical results presented in this section use the BP-RNN model from Section~\ref{ref:BP-RNN}, with the data-pass layer defined in~(\ref{var2check_BP_RNN_BP_friendly_imple}). Each specialized BP-RNN is trained independently, using the training set construction method presented in Section~\ref{sec:Training set}. In addition, we also train an unspecialized  BP-RNN decoder, according to the procedure described in~\cite{nachmani2018deep}, to provide a benchmark for the presented numerical results. We use the same \snr for training and testing, thus, all BP-RNNs are trained for each \snr value ranging from $1$\,dB to   $6$\,dB, with a step of either $0.5$\,dB or $1$\,dB. The choice of the maximum number of decoding iterations during the training and the testing phases,  $I_\text{train}$ and $I_\text{test}$ (see Section~\ref{subsec:Number of decoding iterations}), will be discussed in the next section. Finally, we mention that we used the Keras library for training, with training parameters shown in Table~\ref{tab:keras params}.

\begin{table}[!t]
\centering
\caption{Keras Parameters}
\label{tab:keras params}
    \begin{tabular}{|c|c|} 
    \cline{1-2}  
    \multicolumn{1}{|c|}{\textbf{Parameters}} & \textbf{Parameters values}\\  
    \hline 
    \textbf{Optimizer} &  RMSprop \cite{tieleman2012lecture} \\
    \textbf{(Gradient descent)} & (initialized at a learning rate of $10^{-3}$)\\
    \hline 
    \textbf{Epoch number} & 10 \\  
    \hline 
    \textbf{Training batch size}  & 8192 \\  
    \hline 
    \textbf{Number of batches} & 37 to 122 (depending on the \snr) \\
    \hline 
    \end{tabular}
\end{table}

%
%
%

\subsection{Maximum number of decoding iterations for training and testing}
\label{sec:training ite impact}

\begin{figure*}[!t]
\centering
\subfloat[BP-RNN $D_1$, trained on class $7-(5,8,(5,8))$]{\includegraphics[width=.32\linewidth]{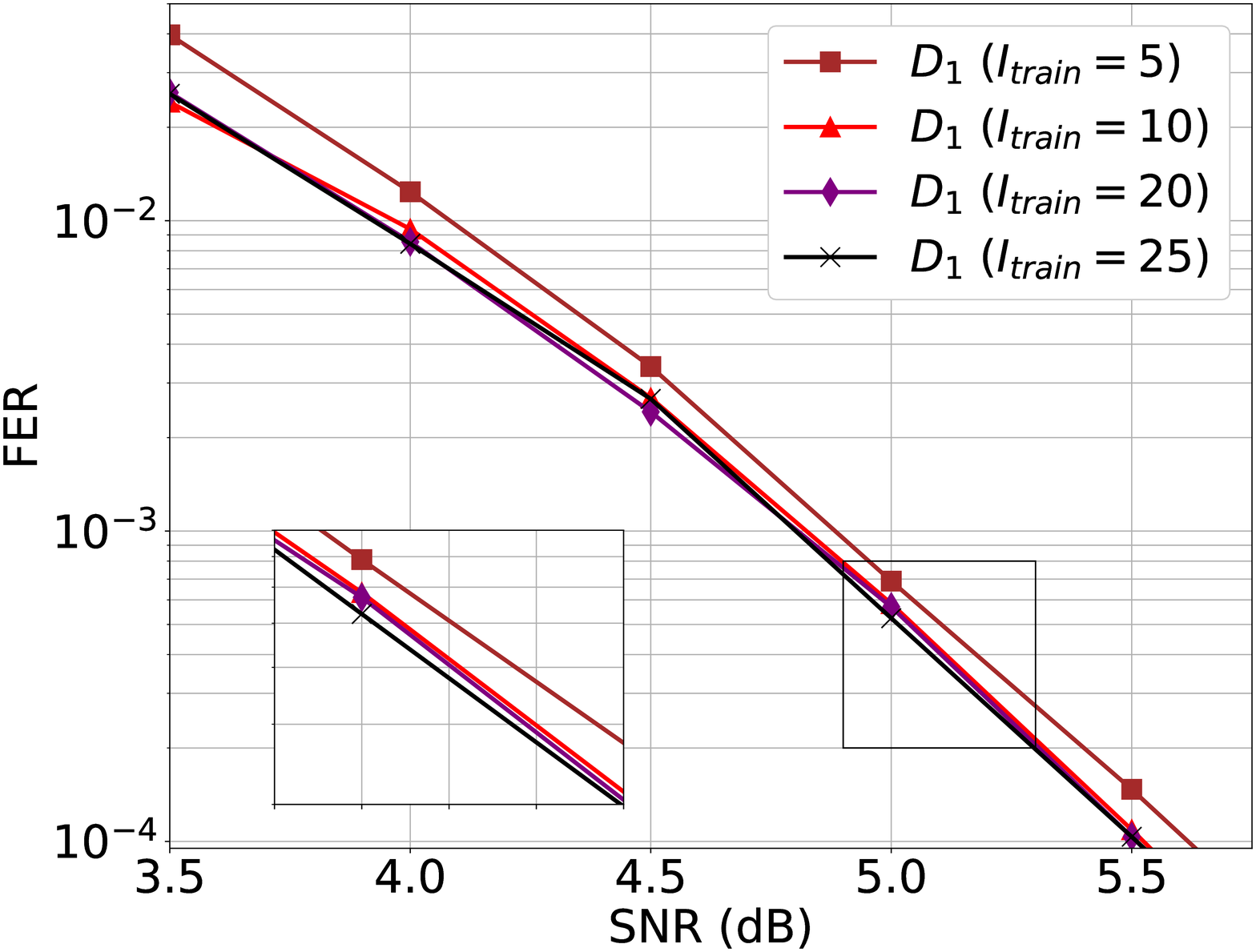}
\label{FER_Code_1_impact_Imaxtrain_BP_RNN_AS}}\quad%
\subfloat[BP-RNN \cite{nachmani2018deep}]{\includegraphics[width=.32\linewidth]{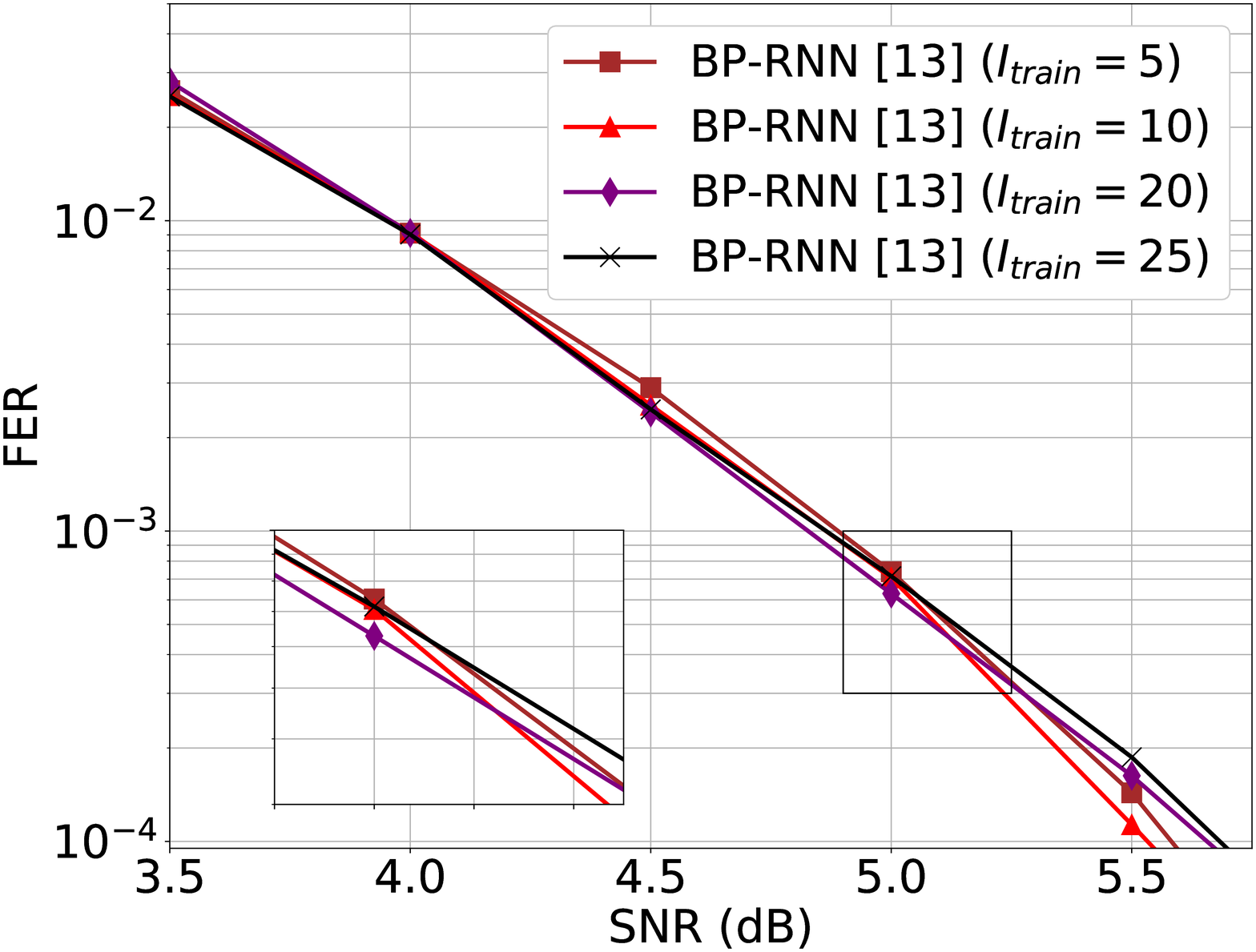}
\label{FER_Code_1_impact_Imaxtrain_BP_RNN_Nachmani}}\hfill%
\subfloat[All specialized BP-RNNs, serial architecture]{\includegraphics[width=.32\linewidth]{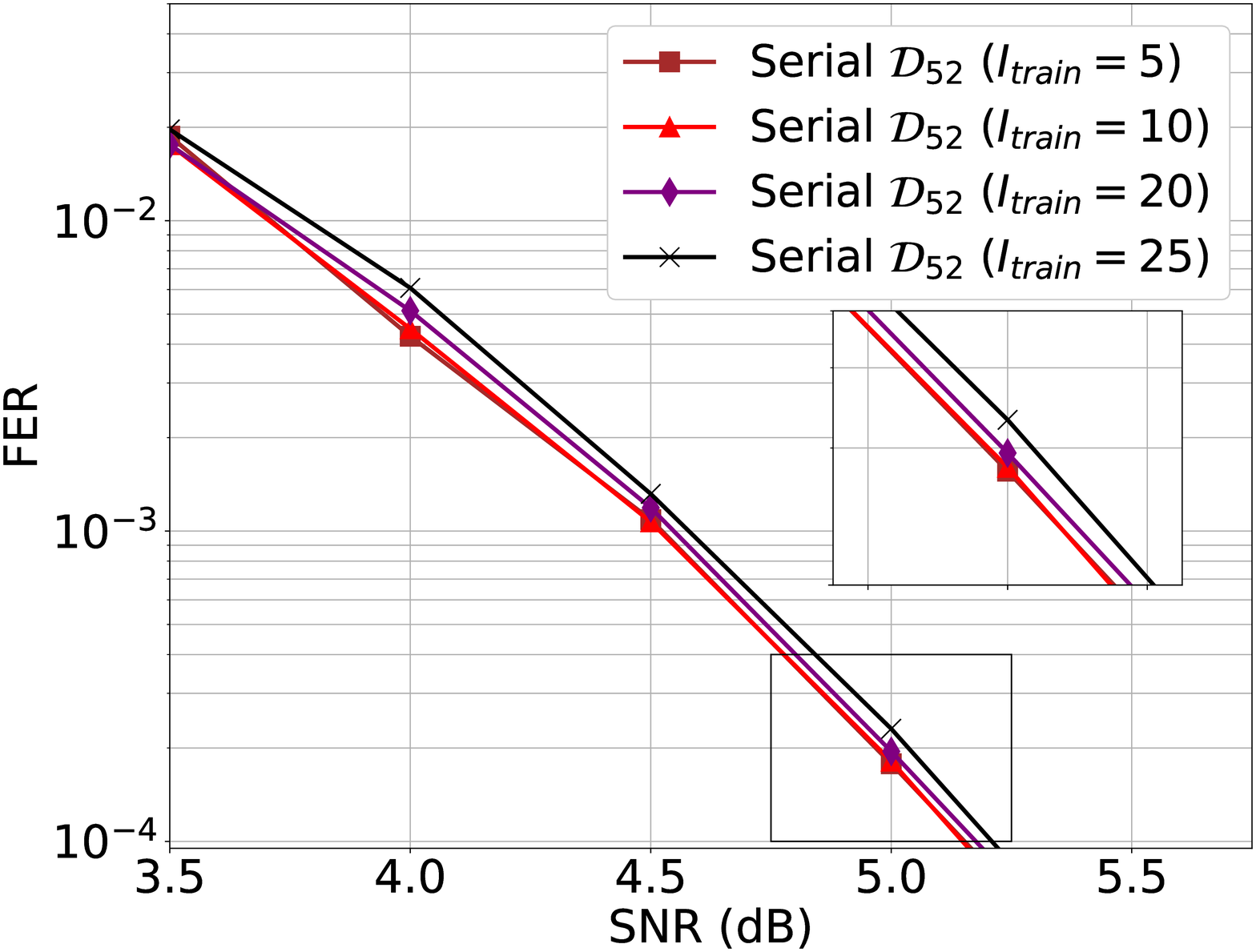}
\label{FER_Code_1_impact_Imaxtrain_serial}}
\caption{Impact of training parameter $I_\text{train}$ on the FER performance, for  BP-RNN decoders using $I_\text{test}=25$ (Code-$1$)}
\label{I_max_train_impact}
\end{figure*}

To illustrate the discussion from Section~\ref{subsec:Number of decoding iterations}, we fix the maximum number of decoding iterations for testing our BP-RNN decoders to $I_\text{test} = 25$ iterations, and investigate here the impact of the maximum number of decoding iterations used at the training phase, $I_\text{train}$.

We consider the Code-1, and train all the BP-RNN decoders for $I_\text{train} \in \left\{5,10,20,25\right\}$. Fig.~\ref{I_max_train_impact} shows the frame error rate (FER) results, using  $I_\text{test} = 25$ iterations, for (a) the  BP-RNN decoder specialized on the error class $7\mhyphen(5,8,(5,8))$, indicated as $D_1$ in the legend\footnote{See also Table~\ref{tab:selected-ET} and the corresponding discussion from Section~\ref{sec:comple results}.}, (b) the unspecialized BP-RNN decoder~\cite{nachmani2018deep}, and (c) the serial architecture using the $52$ specialized BP-RNN decoders.  We observe no noticeable difference on the FER performance, except in Fig.~\ref{FER_Code_1_impact_Imaxtrain_BP_RNN_AS}, where the FER performance for $I_\text{train} = 5$ is slightly degraded with respect to $I_\text{train} \in \left\{10,20,25\right\}$. In the following, we choose $I_\text{train}=10$, which allows for faster training. We can note that similar observations hold for Code-2 (not shown here).

\subsection{BP-RNN diversity selection} 
\label{sec:comple results}

\begin{figure}
    \includegraphics[width=.7\linewidth]{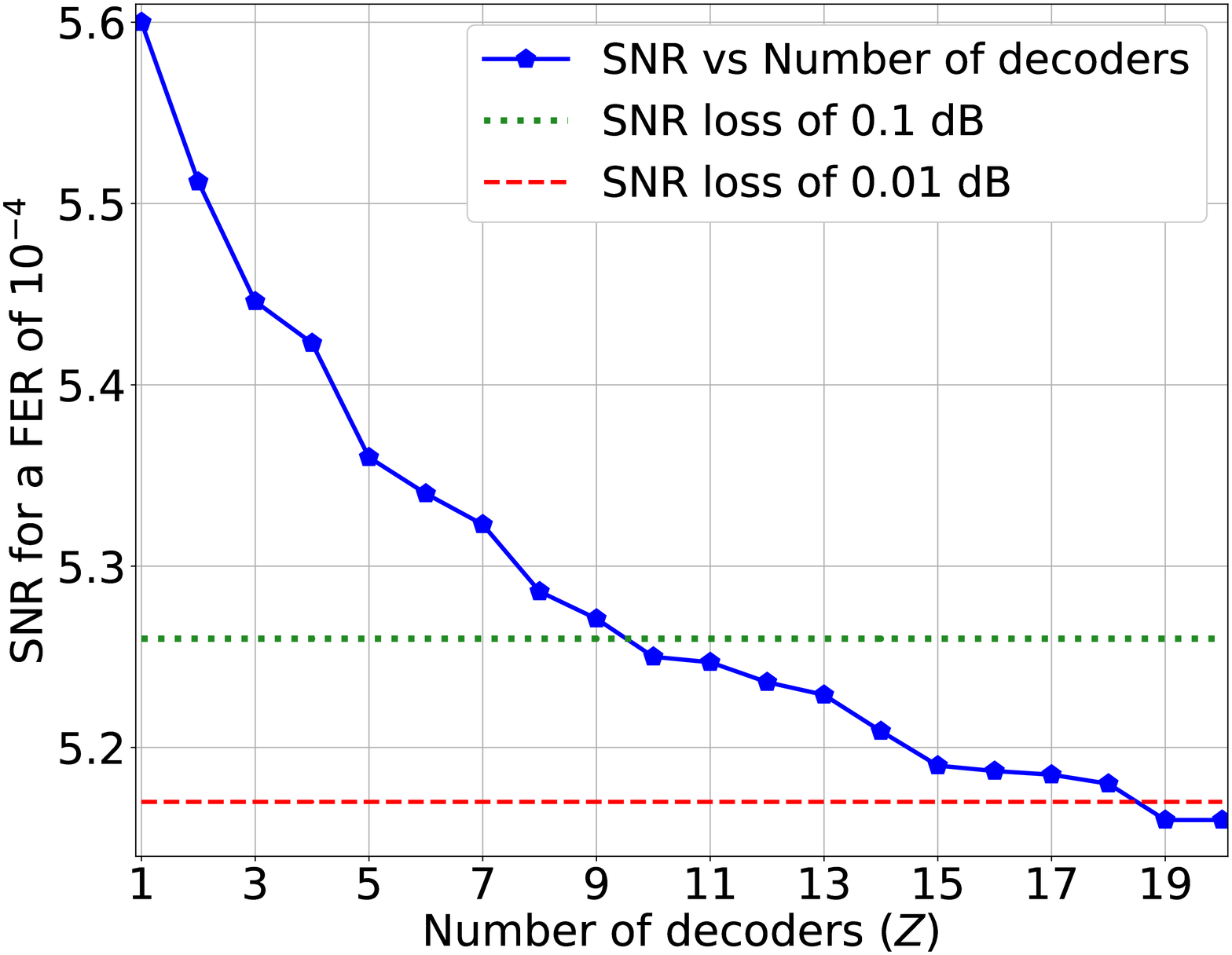}
    \centering
    \caption{\snr for target $\text{FER} = 10^{-4}$, as function of $Z$ (Code-$1$)}
    \label{sub_opti select code 1}
\end{figure}

We apply now the BP-RNN diversity selection procedure from 
Section~\ref{subsec:BP-RNN diversity selection}, to both Code-1 and Code-2. We proceed as follows. First, we fix $\snr = 5$\,dB, and we train $J$ BP-RNN decoders specialized on the $J$ error classes ($J=52$ for Code-1, $J=120$ for Code-2). 
We order these decoders according to the procedure described in Section IV-A, where we use a common test set $T$ containing $10^{8}$ noisy codewords ($\snr = 5$\,dB) to assess their individual error correction performance. Then we select a diversity $\mathcal{D}_{Z}$ of BP-RNN decoders according to~(\ref{eq:diversity_size_Z}), corresponding to $Z$ different error classes.
Subsequently, we only consider BP-RNN decoders specialized on the selected error classes, but we train them again for each \snr value used for testing\footnote{We have also performed simulations using  the BP-RNN decoders trained for $\snr = 5$\,dB only, and testing them on different \snr values. Our simulation results were very similar to those obtained by training again the BP-RNN decoders for the actual \snr value used for testing. We chose to show  simulation results for the case where the training and testing \snr values are equal, simply for consistency reasons.}.

For Code-1, Fig.~\ref{sub_opti select code 1} shows the \snr required for achieving a target $\text{FER} = 10^{-4}$, as a function of the number  of selected BP-RNN decoders $Z$ (only results for $Z=1,\dots,20$ are shown, since we observe no further improvement of the FER for higher $Z$ values). We choose $Z=10$, corresponding to an SNR loss of less than $0.1$\,dB, with respect to the case when all the $J$ BP-RNN decoders are used. We use the same procedure to select a number of BP-RNN decoders for Code-2, which yields the same value $Z=10$, for which the SNR loss is again less than $0.1$\,dB. Finally, in Table~\ref{tab:selected-ET}, we show the extended types for the $Z=10$ error classes, corresponding to the selected decoders.

\begin{table}[!t]
\centering
\caption{Extended Types (ET) for the Selected Error Classes}
\label{tab:selected-ET}
\begin{tabular}{|l|c||l|c|}
\hline
\multicolumn{2}{|c||}{Code-1} & \multicolumn{2}{c|}{Code-2} \\
\hline
 Dec. & Error-Class (ET) & Dec. & Error-Class (ET) \\
 \hline\hline
 $D_1$  & $7\mhyphen(5,8,(5,8))$ & $D_1$ & $7\mhyphen(7,11,(6,11,1)) $ \\ 
 $D_2$  & $7\mhyphen(1,9,(0,9,1))$   & $D_2$ & $5\mhyphen(7,9,(7,9)) $ \\ 
 $D_3$  & $4\mhyphen(2,5,(2,5))$     & $D_3$ & $6\mhyphen(4,10,(4,10)) $ \\ 
 $D_4$  & $7\mhyphen(3,7,(1,7,2))$   & $D_4$ & $7\mhyphen(5,9,(5,9)) $ \\ 
 $D_5$  & $8\mhyphen(2,9,(1,8,1,1))$ & $D_5$ & $6\mhyphen(8,10,(8,10)) $ \\ 
 $D_6$  & $6\mhyphen(2,7,(2,6,0,1))$ & $D_6$ & $7\mhyphen(5,11,(4,11,1)) $ \\ 
 $D_7$  & $5\mhyphen(1,7,(1,7))$     & $D_7$ & $7\mhyphen(5,8,(5,8)) $ \\ 
 $D_8$  & $6\mhyphen(2,7,(1,7,1))$   & $D_8$ & $7\mhyphen(7,7,(7,7)) $ \\ 
 $D_9$  & $7\mhyphen(1,9,(1,8,0,1))$ & $D_9$ & $7\mhyphen(3,11,(3,11)) $ \\ 
 $D_{10}$ & $3\mhyphen(3,3,(3,3))$     & $D_{10}$ & $7\mhyphen(7,13,(7,13))$ \\
 \hline
\end{tabular}
\end{table}

\subsection{FER results, complexity and decoding latency evaluations}
\label{sec:FER results}
\begin{figure*}
\centering
\subfloat[FER results for Code-$1$]{\includegraphics[width=.32\linewidth]{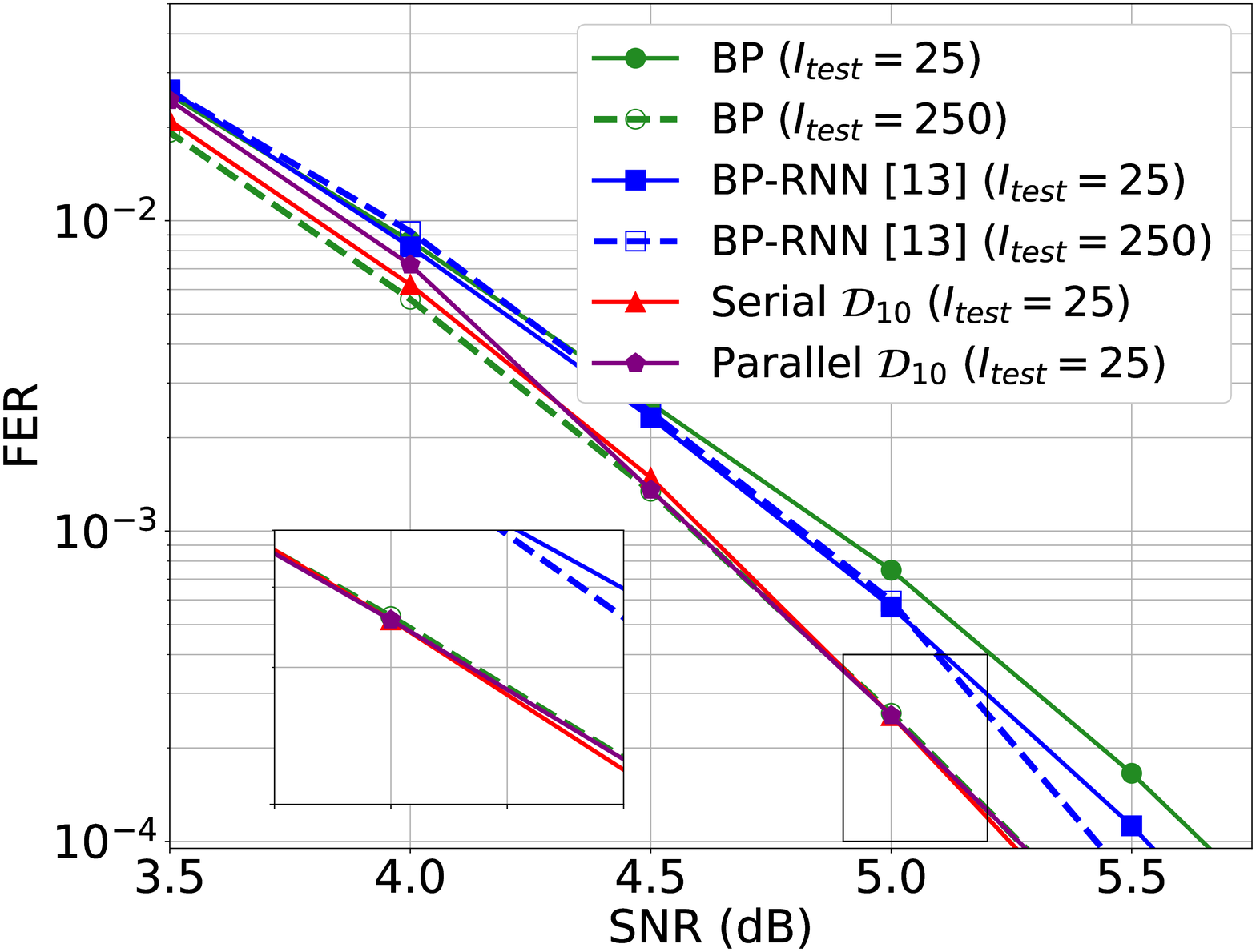}\label{FER_Code_1}}\hfill%
\subfloat[FER results for Code-$2$]{\includegraphics[width=.32\linewidth]{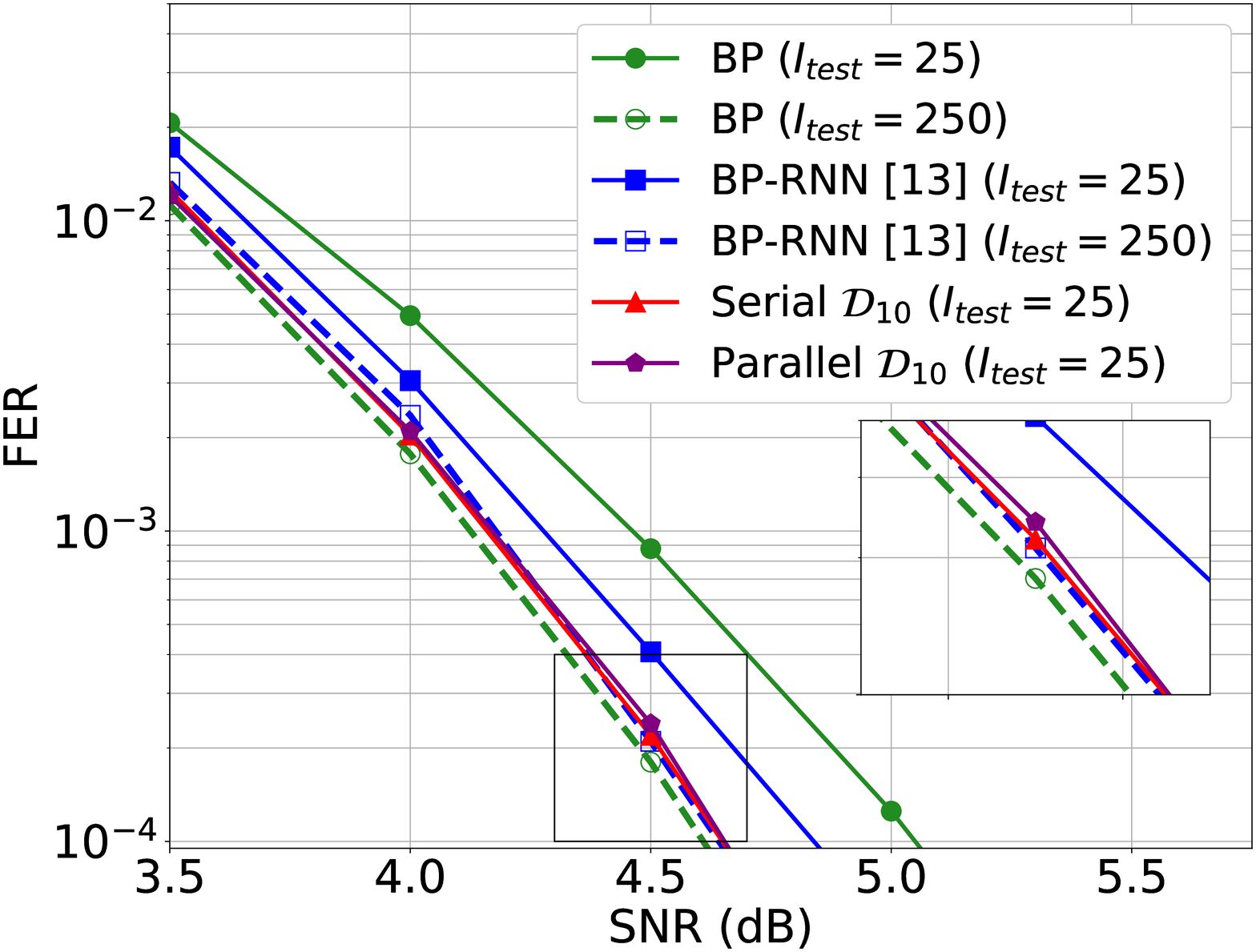}\label{FER_Code_2}}\hfill%
\subfloat[Complexity/decoding latency for Code-$2$]{\includegraphics[width=.32\linewidth]{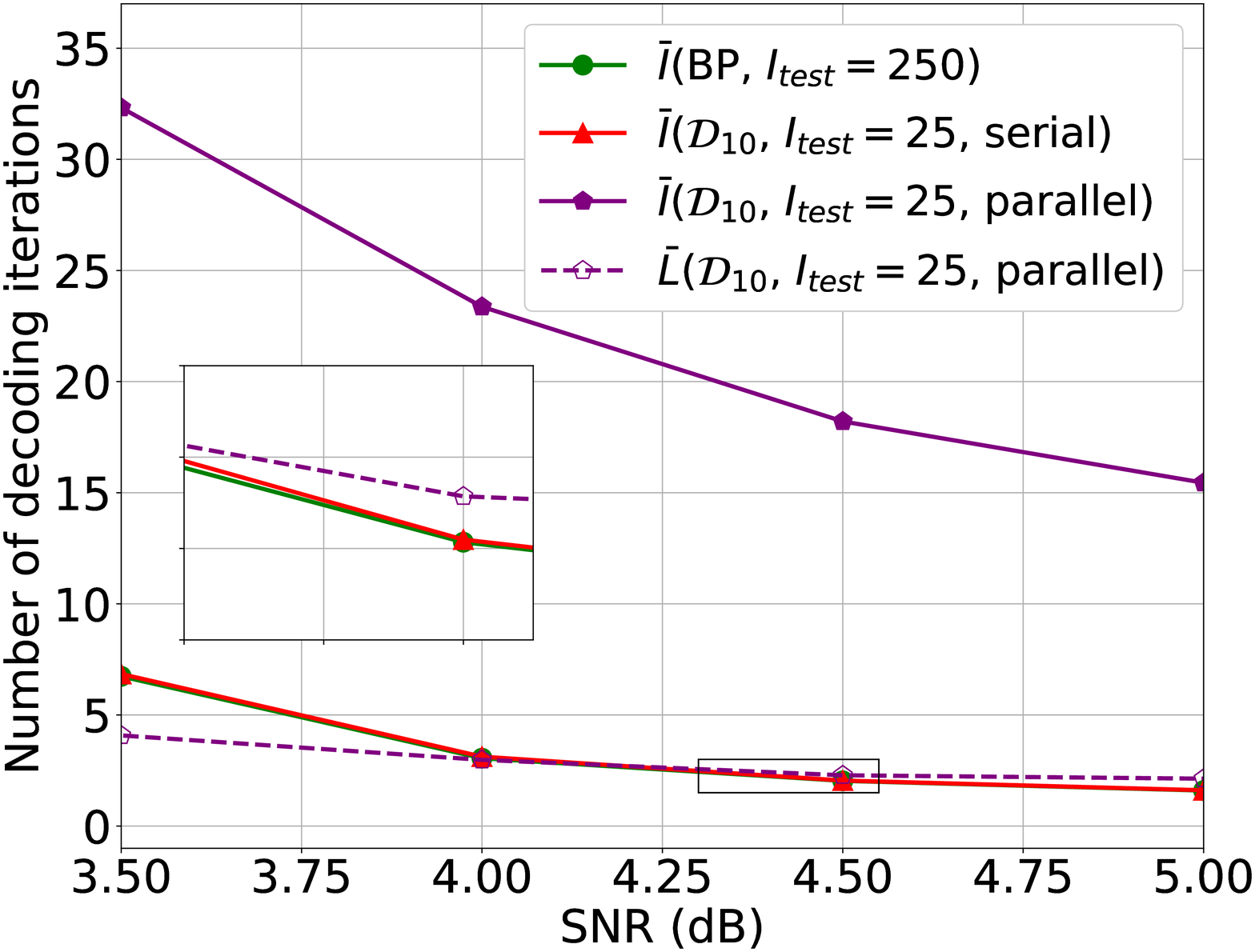}\label{Complexity code 2}}
\caption{FER results, complexity and decoding latency evaluations (without OSD post-processing)}
\label{FER_Code1_Code2}
\end{figure*}

We consider the BP-RNN diversity $\mathcal{D}_{10}$ composed of the $Z=10$ BR-RNN decoders selected in the previous section, and evaluate the FER performance, as well as the average-case complexity and decoding latency, for both parallel and serial architectures from Section~\ref{subsec:bprnn_diversity_architectures}. 

Fig.~\ref{FER_Code1_Code2} shows the FER results for (a) Code-1, and (b) Code-2. For comparison purposes,  we also show the FER performance of the BP decoder and the BP-RNN decoder from~\cite{nachmani2018deep}, with either $I_\text{test} = 25$, or  $I_\text{test} = 250$. 
The latter $I_\text{test}$ value corresponds to the cumulative maximum number of iterations performed by the BP-RNN decoders in the diversity $\mathcal{D}_{10}$.
First, we observe that the parallel and serial architectures exhibit virtually the same FER performance, the corresponding curves being practically superimposed one on another. 
Compared to the conventional BP decoding, the BP-RNN diversity $\mathcal{D}_{10}$ produces an SNR gain of approximately $0.4$\,dB with respect to BP($I_\text{test} = 25$). This comparison is relevant to applications with strict latency requirements, since both the BP($I_\text{test} = 25$) and the parallel BP-RNN diversity have the same worst-case decoding latency. By way of comparison, for Code-2, a similar gain over the conventional BP has been recently reported in~\cite[Fig.~6]{geiselhart2022automorphism}, by using an automorphism ensemble decoding (AED) approach, with $16$ BP decoders working in parallel\footnote{We did not include the AED-16 curve from~\cite{geiselhart2022automorphism} in Fig.~\ref{FER_Code_2}, to avoid clutter. The gain reported in~\cite{geiselhart2022automorphism} was observed using $32$ decoding iterations.}. If the worst-case latency constraint is relaxed, it can be observed from Fig.~\ref{FER_Code1_Code2} that BP($I_\text{test} = 250$) achieves similar FER performance to the BP-RNN diversity $\mathcal{D}_{10}$. Similar considerations hold for the comparison between the (unspecialized) BP-RNN~\cite{nachmani2018deep} and the BP-RNN diversity $\mathcal{D}_{10}$, with one noticeable exception for Code-1, for which it can be observe that increasing the number of iterations from $I_\text{test} = 25$ to $I_\text{test} = 250$ does not improve the FER performance of the BP-RNN~\cite{nachmani2018deep} decoder (or only slightly in the low FER region).
Finally, we note that using only the first decoder ($D_1$) of our specialized BP-RNN decoders yields similar FER performance to the unspecialized  BP-RNN~\cite{nachmani2018deep}. While this is not shown in the figure (to avoid clutter),  it can be observed for Code-1 by comparing the FER results in Figs.~\ref{FER_Code_1_impact_Imaxtrain_BP_RNN_AS} and~\ref{FER_Code_1_impact_Imaxtrain_BP_RNN_Nachmani}.

\begin{figure*}[!th]
\hspace*{-0.32cm}
\centering
\subfloat[Code-$1$]{\includegraphics[width=0.42\linewidth]{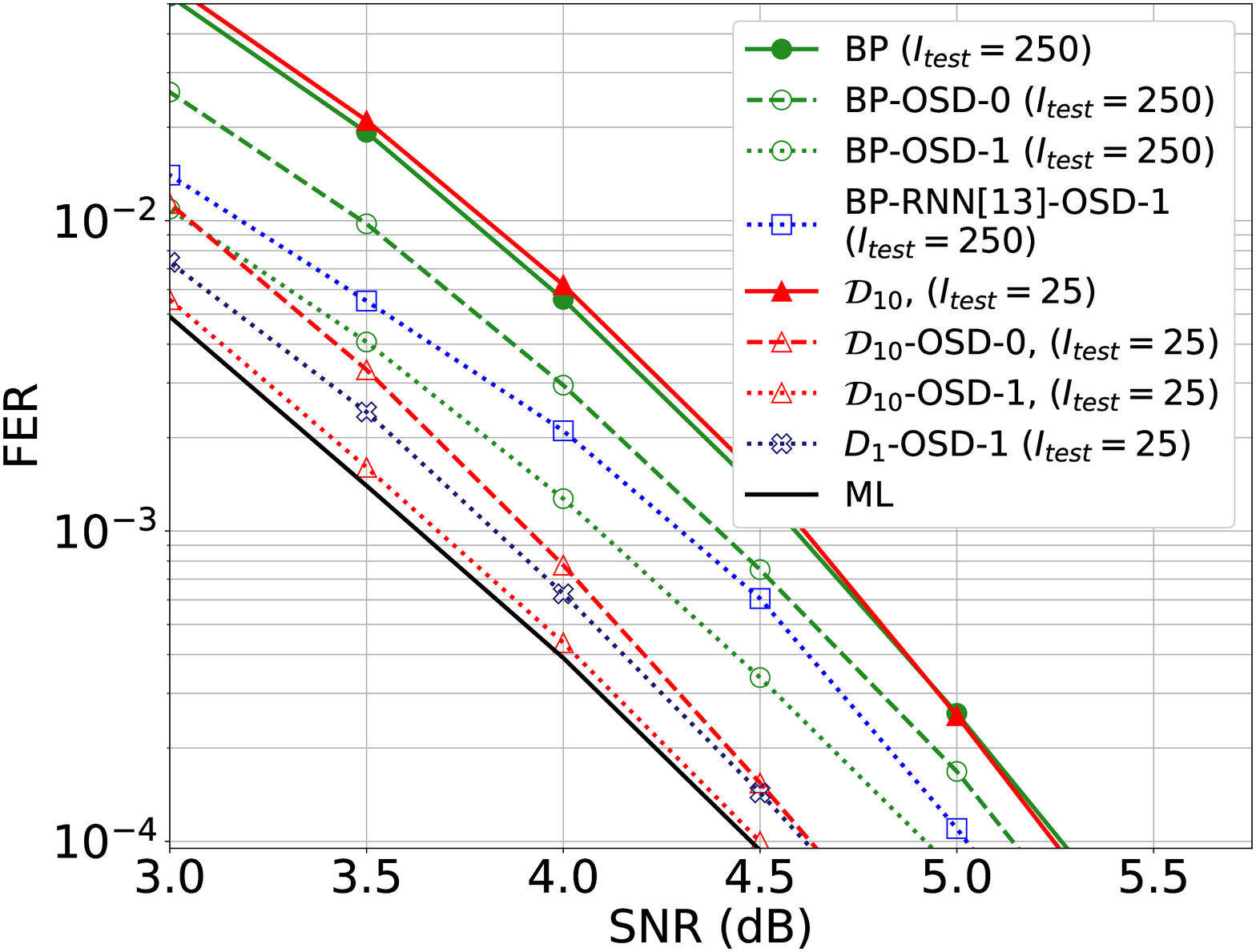}\label{FER_Code_1_OSD}}%
\qquad
\qquad
\subfloat[Code-$2$]{\includegraphics[width=0.42\linewidth]{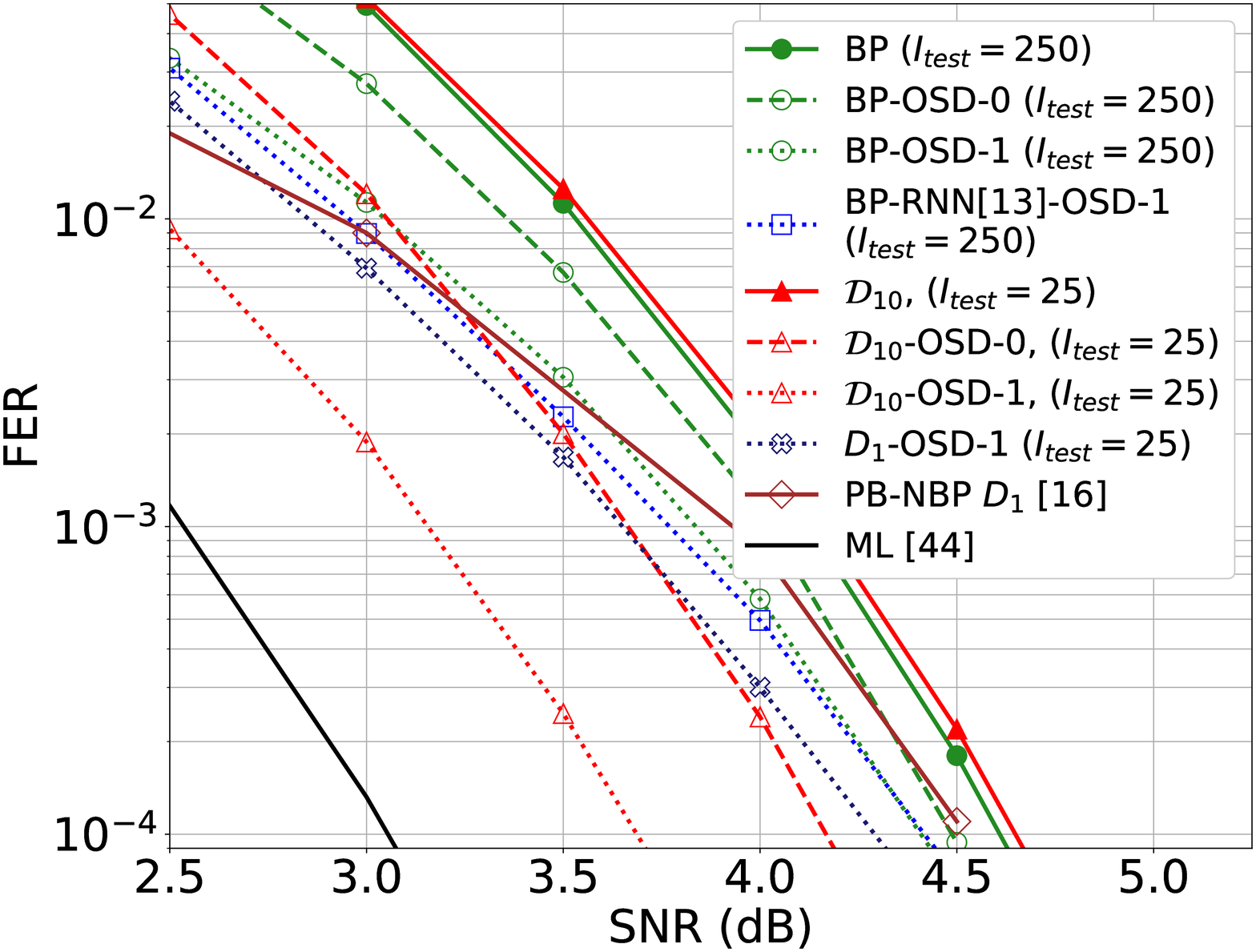}\label{FER_Code_2_OSD}}
\caption{FER results using OSD post-processing}
\label{FER_Code_1_2_OSD}
\end{figure*}

Fig.~\ref{Complexity code 2} shows the average-case computational complexity and the average-case decoding latency results (Code-2 only), evaluated using (\ref{eq:ave_ite_para}) and (\ref{eq:ave_lat}) respectively. We compare the BP($I_\text{test}=250$) and the BP-RNN diversity $\mathcal{D}_{10}$ with  either the serial or the parallel architecture, as they achieve similar FER performance. For the BP decoder and the serial $\mathcal{D}_{10}$, the average number of decoding iterations ($\bar{I}$) is a measure of both average-case computational complexity and average-case decoding latency. For the parallel  $\mathcal{D}_{10}$, $\bar{I}$ only measures the average-case computational complexity, while the average-case decoding latency is measured by $\bar{L}$.  While the parallel $\mathcal{D}_{10}$ has the lower worst-case latency, it can be observed that it also exhibits a reduced average-case latency in the first part of the waterfall (up to $4$\,dB).   The serial $\mathcal{D}_{10}$  may be seen as an alternative to BP($I_\text{test} = 250$), providing not only similar decoding performance, but also similar worst- and average-case computational complexity and decoding latency. Yet, both the parallel and the serial $\mathcal{D}_{10}$ retain the advantage of decoding diversity, which can be conveniently  exploited by the proposed OSD post-processing step, as illustrated in the next section.

\subsection{FER results using OSD post-processing}
\label{sec:OSD results}

We consider low-order ($w=0,1$) OSD post-processing applied to  the BP($I_\text{test}=250$), the unspecialized BP-RNN($I_\text{test}=250$) decoder~\cite{nachmani2018deep}, and the BP-RNN diversity $\mathcal{D}_{10}(I_\text{test}=25)$. Since  the parallel and the serial $\mathcal{D}_{10}$ exhibit similar FER performance, and  OSD post-processing is only applied in case all the BP-RNN decoders composing $\mathcal{D}_{10}$ fail to find a codeword,  it follows that the OSD post-processing step yields similar  performance when applied to either one of the parallel or serial architecture. We simply refer to the corresponding decoder as $\mathcal{D}_{10}$-OSD, without mention of the diversity architecture.

Simulation results are presented in Fig.~\ref{FER_Code_1_2_OSD} for (a) Code-1, and (b) Code-2. For Code-1, first we note that the BP-OSD-1 provides better performance than the unspecialized BP-RNN-OSD-1. Using only the first decoder of our BP-RNN diversity ($D_1$-OSD-1) outperforms the BP-OSD-1 by about $0.31$\,dB at $\text{FER}=10^{-4}$. Using all the BP-RNN diversity ($\mathcal{D}_{10}$-OSD-1) provides an extra gain of $0.12$\,dB, \emph{i.e.}, a total gain of about $0.43$\,dB with respect to BP-OSD-1. Furthermore, we observe that $\mathcal{D}_{10}$-OSD-1 virtually achieves the ML decoding performance, where the latter is estimated according to~\cite{fossorier1995osd} (we also note that the OSD-3 decoder provides an accurate approximation of the ML decoding performance). Finally, a gain of $0.52$\,dB can be observed for $\mathcal{D}_{10}$-OSD-0 with respect to BP-OSD-0. 

For Code-2,  we note that the unspecialized BP-RNN-OSD-1 provides slightly better performance than the BP-OSD-1. $\mathcal{D}_{10}$-OSD-$w$ outperforms $\text{BP}$-OSD-$w$, by $0.32$\,dB, for $w=0$, and $0.72$\,dB, for $w=1$, at $\text{FER}=10^{-4}$. Using only the first decoder of our BP-RNN diversity, we observe that $D_1$-OSD-1 outperforms BP-OSD-1 by about $0.11$\,dB.
For comparison purposes, we have also included in Fig.~\ref{FER_Code_2_OSD} the FER performance of the Pruning Based Neural BP (PB-NBP) decoder from~\cite{buchberger2020pruning}. Several PB-NBP variants are presented in ~\cite{buchberger2020pruning}, we consider here the PB-NBP decoder $D_1$ (see Fig.~6 in \emph{loc. cit.}). It can be observed that $\mathcal{D}_{10}$-OSD-$1$ outperforms the PB-NBP decoder by $0.84$\,dB, our decoder achieving a FER performance at only $0.63$\,dB from the ML decoding.~\nocite{channelcodes}

To further analyze the diversity obtained by $\mathcal{D}_{10}$-OSD-$w$, we compare it with a decoding diversity combining BP and OSD, where the latter is applied periodically throughout the  BP decoding iterations~\cite{fossorier2001iterative}. Precisely, we consider BP($I_{test}=250$), and record the a-posteriori LLR values after every 25 decoding iterations. In case a codeword is not found after 250 decoding iterations (BP fails), we apply one OSD-$w$ on each recorded set of a-posteriori values. Hence, the OSD-$w$ is carried out ten times in case of a decoding failure, as with $\mathcal{D}_{10}$-OSD-$w$. The ML rule~(\ref{ML rule}) is used afterwards to determine the final codeword.
\begin{figure}[!t]
\includegraphics[width=0.86\linewidth]{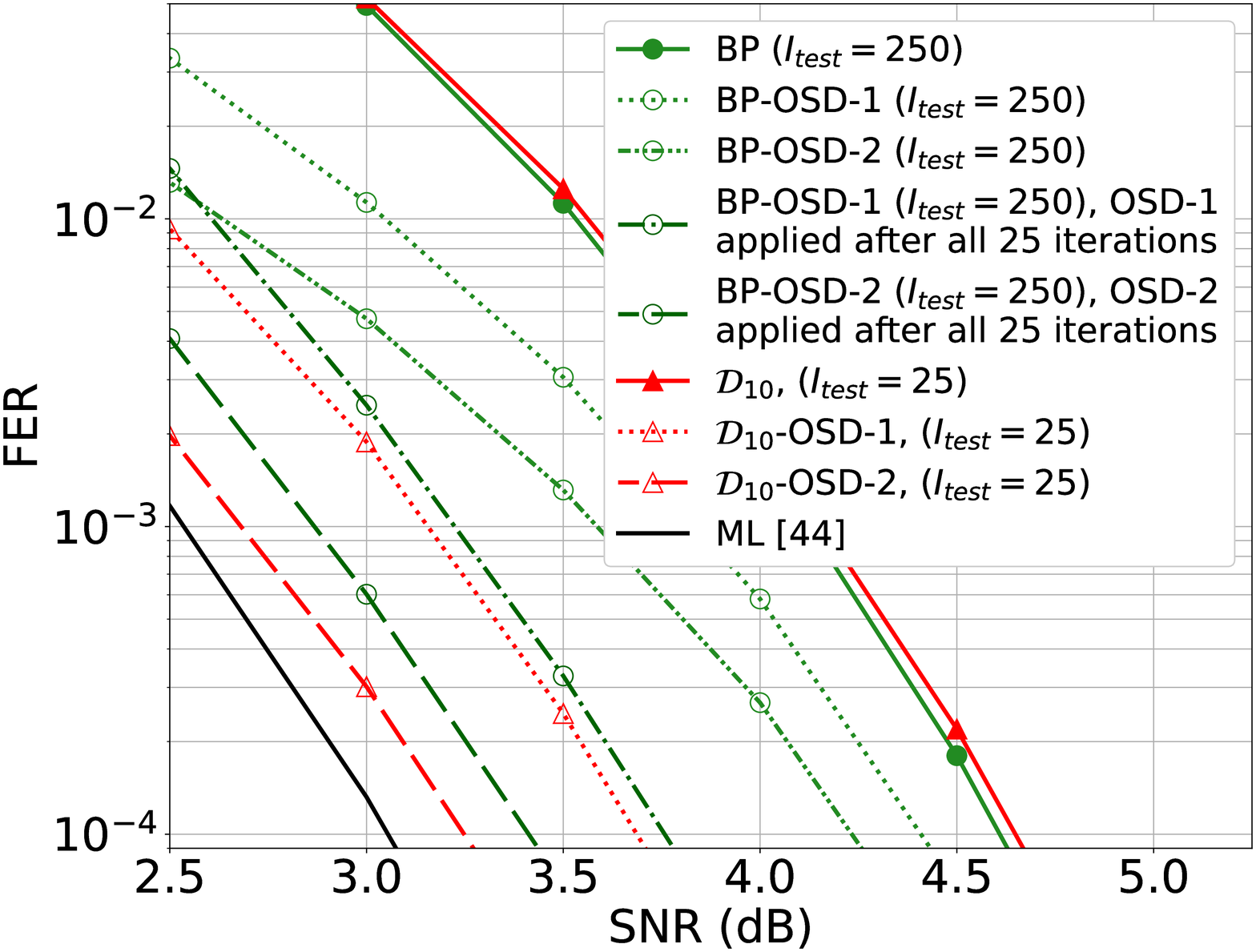}
\centering
\caption{\textnormal{FER results using OSD post-processing of order $w=1$ and $w=2$ for Code-2}}
\label{FER_OSD_2}
\end{figure}

We restrict our attention to Code-2 and, in order to further narrow the gap to ML decoding performance, we consider OSD of order $w=1$ and $w=2$. The corresponding FER results are shown in Fig.~\ref{FER_OSD_2}.  We observe that the proposed $\mathcal{D}_{10}$-OSD-$w$ performs better than BP($I_{test}=250$) with OSD-$w$ every 25 iterations, especially for $w=2$, where it achieves a FER performance within $0.2$\,dB from the ML decoding. Moreover, it should be noticed that using the parallel decoding architecture from Section IV.B, the worst case latency of the $\mathcal{D}_{10}$ diversity corresponds to 25 decoding iterations, while the worst case latency of BP($I_{test}=250$) is equal to 250 decoding iterations.  Finally, we note that $\mathcal{D}_{10}$-OSD-2 exhibits a gain of $1$\,dB with respect to BP($I_{test}=250$)-OSD-2.

\section{Conclusion}
\label{sec:conclusion}
We addressed the problem of enhancing the BP-RNN performance at short code length, by exploring two complementary approaches: (1) decoding diversity, derived by specializing BP-RNN decoders to specific classes of errors, and (2) reliability-based post-processing. We showed that the  proposed BP-RNN diversity coupled with a parallel decoding architecture allows increasing the error correction capability, without increasing the worst-case latency. Moreover, we showed that the proposed OSD post-processing step advantageously  leverage the bit-error rate optimization induced by the use of the cross-entropy loss function, as well as the diversity brought by the use of multiple BP-RNN decoders. The proposed approach, combining decoding diversity and low-order OSD post-processing, provides an efficient way to bridge the gap to ML decoding. It also opens new perspectives for the emerging domain of NN-based decoders. Indeed, we believe that new approaches may be considered for the optimization of NN-based decoders, not to deliver the best possible bit or frame error rate performance, but merely an output that best suits the reliability-based post-processing step.

\section*{Availability of Data}

Parity check-matrices of Code-1 and Code-2, as well as the complete list of absorbing sets of size $\nu \leq 8$, are openly available at \url{https://ai4code.projects.labsticc.fr/software/}.

\appendix
\kern-0.5em
\section*{Extended complexity discussion}

In this appendix, we extend the complexity discussion of Section~\ref{sec:FER results}. First, we compare BP-RNN and BP in terms of memory requirements and number of operations. Since our BP-RNN decoders use the data-pass layer defined in~(\ref{var2check_BP_RNN_BP_friendly_imple}), they can be implemented using conventional BP decoding architectures. In this case, the extra memory cost for a single BP-RNN decoder with respect to the BP decoder comes  from the weights storage only, and the BP-RNN decoder also requires an extra multiplication for each weight in~(\ref{apost_BP_RNN}) and~(\ref{var2check_BP_RNN_BP_friendly_imple}).

\def\cent#1{\,\hfill #1 \hfill\,}
\begin{table*}[!t]
\centering

\caption{Decoding complexity (in terms of the number of check-node updates) for Code-2}
\label{tab:complexity-ext}
    \begin{tabular}{|c|p{25mm}|p{25mm}|p{25mm}|c|} 
    \cline{1-5}  
    \multicolumn{1}{|c|}{\raisebox{-.5ex}{\textbf{Decoding}}} & 
    \multicolumn{3}{c|}{\textbf{Decoding complexity (number of check-node updates~\cite{smith2010design})}} &
    \raisebox{-.5ex}{\textbf{Number}} \\ 
    \cline{2-4}  
    \raisebox{.5ex}{\textbf{architecture}} & \cent{\textbf{Worst-case}} & \cent{\textbf{Best-case}} & \cent{\textbf{Average-case$^{(*)}$}} & \raisebox{.5ex}{\textbf{of weights}}\\
    \hline 
    BP($I_{test}=25$)  &  \cent{12800}  & \cent{512} & \cent{983} & 0 \\
    \hline
    BP($I_{test}=250$)  &  \cent{128000} & \cent{512} & \cent{1044} & 0 \\
    \hline
    PB-NBP $D_1$~[16] &  \cent{25920} & \cent{5120} & \cent{N/A} & 28416 \\
    \hline
    Serial $\mathcal{D}_{10}$  & \cent{128000} & \cent{512} & \cent{1044} & 10240  \\
    \hline 
    \multicolumn{5}{l}{$^{(*)}$\,Average-case decoding complexity is evaluated at $\snr = 4.5$\,dB.}
    \end{tabular}
\end{table*}

Since the check-node updates are the most computationally intensive operations in BP decoding, counting their number is also relevant to evaluate the complexity of a BP-based decoder~\cite{smith2010design}. In~\cite{buchberger2020pruning},  the worst case complexity of the proposed PB-NBP decoder is evaluated by counting the number of check-node updates when all decoding iterations are performed. In our case, the number of check-node updates is simply given by the number of edges of the Tanner graph, multiplied by the number of  decoding iterations. We can then evaluate the worst-case decoding complexity (corresponding to the maximum number of decoding iterations), the best-case complexity (corresponding to one decoding iteration only), and the average-case complexity (corresponding to the average number of decoding iterations).
The corresponding results for the BP, the PB-NBP\,$D_1$~\cite{buchberger2020pruning}, and the serial $\mathcal{D}_{10}$ decoders are given in Table~\ref{tab:complexity-ext}. In the worst-case scenario, we observe that the serial $\mathcal{D}_{10}$ has the same decoding complexity as BP($I_{test}=250$), which is 4.85 times higher than that of PB-NBP\,$D_1$. In the best-case  scenario, the PB-NBP\,$D_1$ decoder is penalized by the fact that it uses a high number of check-nodes during the first iteration (some of which may be punctured during subsequent iterations). Precisely, its first iteration comprises 640 check nodes of degree at least  8 (see~\cite[Fig.~4b]{buchberger2020pruning}), giving a number of check-node updates  greater than or equal to 5120. This amounts to 10 times the best-case decoding complexity (or equivalently, the complexity of 10 decoding iterations) of the serial $\mathcal{D}_{10}$ or the BP decoder. In Fig.~\ref{Complexity code 2}, we notice that the average number of decoding iterations $\bar{I}$ is equal to 10 for an $\snr = 3.37$\,dB for both BP and serial $\mathcal{D}_{10}$. As a result, we conclude the serial $\mathcal{D}_{10}$ or BP is less complex than PB-NBP\,$D_1$ in the waterfall region. To further illustrate this, we observe in Table~\ref{tab:complexity-ext} that the serial $\mathcal{D}_{10}$ decoder has an average decoding complexity at $\snr = 4.5$\,dB that is 5 times lower than the best-case decoding complexity of PB-NBP\,$D_1$.
Finally, we note that PB-NBP\,$D_1$ has 2.75 times more weights than $\mathcal{D}_{10}$, due to its neural network architecture, requiring therefore a higher weight storage cost.

\bibliographystyle{IEEEtran}
\bibliography{biblio}
\end{document}